\documentclass{aa}

\usepackage{graphicx}
\usepackage{lscape}
\usepackage{txfonts}
\begin{document}

   \title{Metallicities and ages for 35 star clusters and their
   surrounding fields in the Small Magellanic Cloud}
   \titlerunning{Metallicities and ages for 35 star clusters and their
   surrounding fields in the Small Magellanic Cloud}

   \author{W. Narloch\inst{1}\fnmsep\thanks{contact author},
          G. Pietrzy\'nski\inst{1,2}, W. Gieren\inst{1}, A.~E. Piatti\inst{3,4},
          M. G\'orski\inst{1,2}, P. Karczmarek\inst{1}, D. Graczyk\inst{5}, \\
          K. Suchomska\inst{2}, B. Zgirski\inst{2}, P. Wielg\'orski\inst{2},
          B. Pilecki\inst{2}, M. Taormina\inst{2}, M. Ka\l uszy\'nski\inst{2},
          W. Pych\inst{2}, G. Hajdu\inst{2} \\
           \and
          G. Rojas Garc\'ia\inst{2}
          }
   \authorrunning{W. Narloch et al.}

   \institute{Univesidad de Concepci\'on, Departamento de Astronomia,
              Casilla 160-C, Concepci\'on, Chile\\
              \email{wnarloch@astro-udec.cl}
         \and
             Nicolaus Copernicus Astronomical Center, Polish Academy of Sciences,
              Bartycka 18, 00-716, Warsaw, Poland
         \and
            Instituto Interdisciplinario de Ciencias B\'asicas (ICB), CONICET-UNCUYO,
            Padre~J. Contreras 1300, M5502JMA, Mendoza, Argentina
         \and
            Consejo Nacional de Investigaciones Cient\'{\i}ficas y T\'ecnicas (CONICET),
            Godoy Cruz 2290, C1425FQB,  Buenos Aires, Argentina
         \and
            Nicolaus Copernicus Astronomical Center, Polish Academy of Sciences,
            Rabia\'nska 8, 87-100 Toru\'n, Poland
             }

   \date{Received ; Accepted 14 January 2021}


  \abstract
   {}
   {In this work we study 35 stellar clusters in the Small Magellanic Cloud (SMC)
   in order to provide their mean metallicities and ages.
   We also provide mean metallicities of the fields surrounding the clusters.}
   {We used Str\"omgren photometry obtained with the 4.1 m SOAR telescope and take
   advantage of $(b-y)$ and $m1$ colors for which there is a~metallicity
   calibration presented in the literature.}
   {The spatial metallicity and age distributions of clusters across the SMC
   are investigated using the results obtained by Str\"omgren photometry. We confirm
   earlier observations that younger, more metal-rich star clusters are concentrated
   in the central regions of the galaxy, while older, more metal-poor clusters are located farther
   from the SMC center. We construct the age--metallicity relation for the studied clusters
   and find good agreement with theoretical models of chemical enrichment, and with other literature
   age and metallicity values for those clusters.
   We also provide the mean metallicities for old and young populations of the field stars
   surrounding the clusters, and find the latter to be  in ~good agreement with recent studies
   of the SMC Cepheid population.
   Finally, the Str\"omgren photometry obtained for this study is made publicly
   available.}
   {}


   \maketitle
%

\section{Introduction}

   The Small Magellanic Cloud (SMC) is a~dwarf irregular galaxy; together
   with the Large Magellanic Cloud (LMC) it forms a~pair of interacting satellites
   of the Milky Way. Because of its proximity, it is an ideal environment for
   various astrophysical studies, of which the chemical evolution is one of the
   most crucial. Clusters serve as tracers of the chemical evolution of the galaxies.
   The derived metallicities and ages of stellar clusters generally follow the
   age--metallicity relation (AMR) of a~given galaxy, which allows us to follow
   the chemical enrichment process of the environment and draw conclusions about
   the galactic history. This makes the AMR an~important tool for understanding the
   chemical evolution of the galaxies.

   There are several methods for determining stellar metallicities.
   Spectroscopy is a~very good tool for obtaining the metallicities of stars;
   the best for this purpose are high-resolution spectra with wide spectral range
   and high signal-to-noise ratio. However, it is challenging to get good-quality
   high-resolution spectra of stars from nearby galaxies.  An alternative is the
   use of  low-resolution spectra and the Ca~II triplet (CaT).
   There have been several spectroscopic studies of the SMC star clusters in the
   literature based on CaT. \citet{DH1998} obtained spectra of individual
   red-giant-branch (RGB) stars in seven SMC clusters and calculated the mean
   metallicities for six of them.
   They found abnormally low metallicities for Lindsay~113 and NGC~339, suggesting
   that they have different origins, possibly being formed from infalling unenriched
   gas, in contrast to the rest of their studied clusters. \citet{Carrera2008}
   determined metallicities of over 350 RGB stars in 13 fields distributed across
   the SMC, and for the first time  found a~spatial metallicity
   gradient in this galaxy. The average metallicity of the innermost fields was about
   $-1$~dex on the \citet{CG97} metallicity scale (hereafter the  CG97 scale), and
   decreased when moving toward the outermost regions. They related the observed
   metallicity gradient with the age gradient because the youngest, most
   metal-rich stars were concentrated in the central region of the SMC.
   \citet{Parisi2009} calculated metallicities of 270 individual RGB stars inside
   and around 16 SMC clusters. They found a~mean metallicity for their CaT sample of
   $-0.94$~dex on the CG97 scale. Furthermore, they also found that the mean age and
   metallicity of clusters older than 3~Gyr are
   5.8~Gyr and $-1.08$~dex, while for clusters younger than 3~Gyr these values  were 1.6~Gyr
   and $-0.85$~dex, respectively, thus confirming the previous findings. They
   also refuted the hypothesis of \citet{DH1998} regarding the anomalous nature of
   Lindsay~113 and NGC~339. \citet{Parisi2014} further improved the AMR from
   their first work using more accurate photometry obtained for clusters from their
   sample and utilized it to determine the cluster ages.

   \citet{Mighell1998} used archival Hubble Space Telescope (HST) data to study the
   color--magnitude diagrams (CMDs) of seven SMC clusters.
   For this purpose they applied two methods, and adopted weighted mean metallicities
   from both approaches.
   The first technique was the simultaneous reddening and metallicity  (SRM) method,
   which takes as input the magnitude of the horizontal branch (HB), the color of the
   RGB at the level of HB, the shape described by either a~quadratic relation or
   higher order polynomial, and the position of the RGB. As a~result, the SRM
   provides simultaneous metallicity and reddening determination.
   The second method uses the fact that the RGB slope steepens with decreasing metallicity.
   This dependency can be calibrated for specific colors. \citet{Mighell1998} present
   this calibration for V versus  (B-V), while \citet{Mucciarelli2009} presented
   these relations for near-infrared $JHK$ bands in four SMC clusters. This method
   returns metallicity for a~given reddening.

   The filters $C$ and $T_1$ of the Washington photometric system are very effective
   for metallicity and age studies \citep{Piatti2005, Piatti2007a, Piatti2007b,
   Piatti2011, Piatti2012}.
   The difference in $T_1$ magnitude between the red clump (RC) stars and the main
   sequence turn-off point (MSTO) allows us to determine the ages of  stellar populations.
   Metallicity can be estimated by comparing the shape of the RGBs of stellar clusters
   with published standard fiducial globular cluster RGBs. However, this technique
   requires an age-dependent correction to metallicities derived for intermediate-age
   objects, which is the case for the majority of SMC clusters \citep{Parisi2009}.

   Metallicity values of stellar clusters can be also obtained directly by fitting
   theoretical isochrones to the CMDs. For example, \citet{Perren2017} developed the
   Automated Stellar Cluster Analysis (ASteCA) package, which calculates the
   synthetic CMD that best matches the observed cluster CMD for a given set of
   fundamental parameters (metallicity, age, distance modulus, reddening, and mass).

   Metallicity also can be calculated based on calibration between [Fe/H], $(b-y)$
   and $m1=(v-b)-(b-y)$ Str\"omgren colors \citep{Hilker2000,Dirsch2000}.
   This relation is well defined for red stars within a~certain range of $(b-y)$ colors.
   The main advantage of this method is that  metallicities can be obtained for
   a~large number of individual stars. An early successful application of this
   method was performed by \citet{GR1992}, among others, for studies of NGC~330.
   The relation used in that work was later extended  toward lower metallicities
   by \citet{Hilker2000}, and adopted by \citet{Dirsch2000} for studies of six LMC
   clusters and their    surrounding fields obtained with 1.54~m Danish Telescope
   placed in La Silla Observatory, Chile. \citet{Calamida2007} introduced a~calibration
   for red giant stars from old Galactic globular clusters based on $(v-y)$ and $(u-y)$
   colors, which have stronger sensitivity to effective temperature than $(b-y)$.
   \citet{Livanou2013} presented metallicity and age determinations for 15 LMC and
   8 SMC star clusters based on the Str\"omgren data from the Danish Telescope.
   More recently, \citet{Piatti2018} employed Str\"omgren photometry from the 4.1~m
   SOAR telescope (Cerro Pachón, Chile) to investigate four SMC   intermediate-age
   clusters, in their search for  hints of multiple stellar populations. The work of
   \citet{Piatti2019} presents derived metallicities of yellow and red supergiants
   in nine young LMC and four SMC clusters. In a~recent study, \citet{Piatti2020}
   used metallicity calibration for $(v-y) - m1$ colors from \citet{Calamida2007}
   and showed the age effect on Str\"omgren metallicities, in the sense that younger
   clusters appear to be more metal poor, and the difference is a~quadratic function.

   Motivated by these  results, we decided to determine metallicities, based on Str\"omgren photometry,
   for stars belonging to 35 star clusters and their surrounding fields from
   the SMC. We also estimated the ages of stellar
   clusters in our sample using theoretical isochrones. We then constructed the AMR,
   which allowed us to trace the chemical evolution of the SMC. We also present
   here photometric measurements for stars from our fields.
   The metallicities of some clusters calculated from data presented in this work
   have already been published \citep[e.g.,][]{Piatti2018,Piatti2019,Piatti2020}.
   We decided to reanalyze them for a~several reasons. In this work we apply the
   metallicity calibration presented by \citet{Hilker2000}, as it is calibrated
   for a~wide range of metallicities, and therefore can be used for variety of stellar
   clusters. To this end, we used direct color--color transformation equations to
   the standard system, which lowers the errors of the coefficients compared to the
   calibration method presented by \citet{Piatti2019}, among others. Moreover, we rephotometrized
   images and standardized two of the chips of the camera separately, which is a~further
   improvement. We also used proper motions to reject foreground stars and
   adopted the new reddening values coming from the recently published reddening
   maps of the Magellanic Clouds \citep{Gorski2020,Skowron2020}, as well as positions
   and sizes of clusters from the updated catalog of \citet{Bica2020}.

   This paper is organized as follows. In Section~\ref{sec:obsred} we describe our
   observations and data reduction pipeline, as well as the selection of stars used
   for cluster and field metallicity calculation, the adopted reddenings, the metallicity
   determination procedure based on the two-color Str\"omgren diagram, and the age estimation
   procedure.
   In Section~\ref{sec:results} we present the  results obtained for the distribution
   of the metallicities of clusters and their surrounding fields in the SMC, as well
   as the distribution of the clusters' ages and the resulting AMR. In
   Section~\ref{sec:discussion} we compare our AMR with those found in the
   literature. Finally, in
   Section~\ref{sec:summ} we summarize our results and draw the conclusions of this work.

\section{Observations and data reduction}
\label{sec:obsred}
%
   The optical images of fields containing star clusters in the SMC in three
   Str\"omgren filters ($v$, $b$ and $y$) were collected within the Araucaria Project
   \citep{GPB2005} during six nights on 4.1 m Southern Astrophysical Research
   Telescope (SOAR) placed in Cerro Pach\'on in Chile, equipped with the SOAR Optical Imager
   (SOI) camera (program ID: SO2008B-0917, PI: Pietrzy\'nski). Observing nights were
   divided into two runs. The first was on 17, 18, and 19 December 2008 and the
   second on 16, 17, and 18 January 2009. The
   SOI is a~mosaic camera composed of two E2V 2k$\times$4k CCDs (read by four amplifiers).
   The field of view is 5.26$\times$5.26~arcmin$^2$ at a~pixel scale of
   0.077~arcsec$\cdot$pixel$^{-1}$.
   During observations 2$\times$2 pixel binning was used resulting in a~pixel scale
   of 0.154~arcsec$\cdot$pixel$^{-1}$. We observed 29 fields with star clusters in
   the SMC in total, where fields with NGC~330, NGC~265, and NGC~376 were observed
   twice. Single images were taken in the  air mass range $1.43-1.91$, and the average
   seeing in the three filters was about 1.0~arcsec.
   Table~\ref{tab:smc} summarizes information about our data set.

\begin{figure}
   \centering
   \includegraphics[width=\hsize]{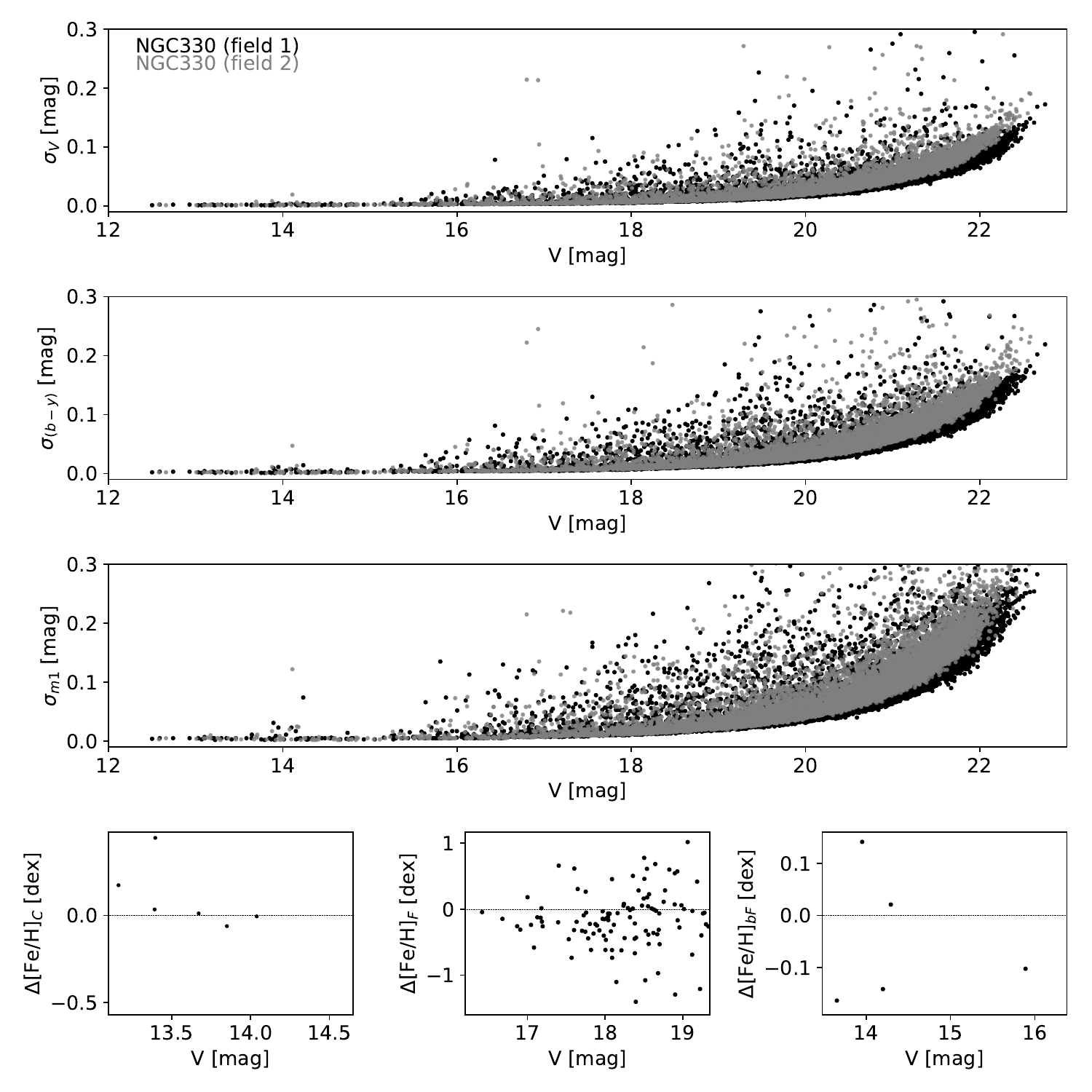}
      \caption{Comparison of photometric precision in two fields of NGC~330
      captured during the nights of 17 and 19 December 2008. Three upper panels:
      photometric errors from DAOPHOT for $V$, $(b-y)$ and $m1$ (black points
      for the field from the first night, gray from the second).
      Three bottom panels: differences in metallicities determined for
      the first and second field for clusters, and old and young field giants, respectively.
      }
      \label{fig:compngc330smc107}
   \end{figure}

   The calibration procedure took into account bias subtraction and flatfield
   correction. Profile photometry was performed with the standard DAOPHOT/ALLSTAR
   package \citep{Stetson1987} using a~Gaussian function with spatial variability
   to define the point spread function (PSF).
   In the case of dense fields, images were additionally divided into smaller
   overlapping subframes to further reduce the PSF and background variability.
   The PSF model was constructed from about 30 to over 200 stars depending on the
   stellar density of a given image. The master list of stars in a~given frame was
   obtained iteratively by gradually decreasing the detection threshold, and in
   the last iteration inspected by eye to  manually add stars omitted in the automatic
   procedure. The aperture corrections for each frame were calculated using the
   DAOGROW package \citep{Stetson1990}, and  instrumental CMDs were
   constructed. The average errors of the photometry were 0.02~mag in $V$ and
   $(b-y)$ and 0.04~mag in $m1$ for stars with brightness $V<20$ mag.
   Figure~\ref{fig:compngc330smc107} illustrates the precision of our photometry
   on an example of the NGC~330 observed during the first and third night.

\begin{figure}
   \centering
   \includegraphics[width=\hsize]{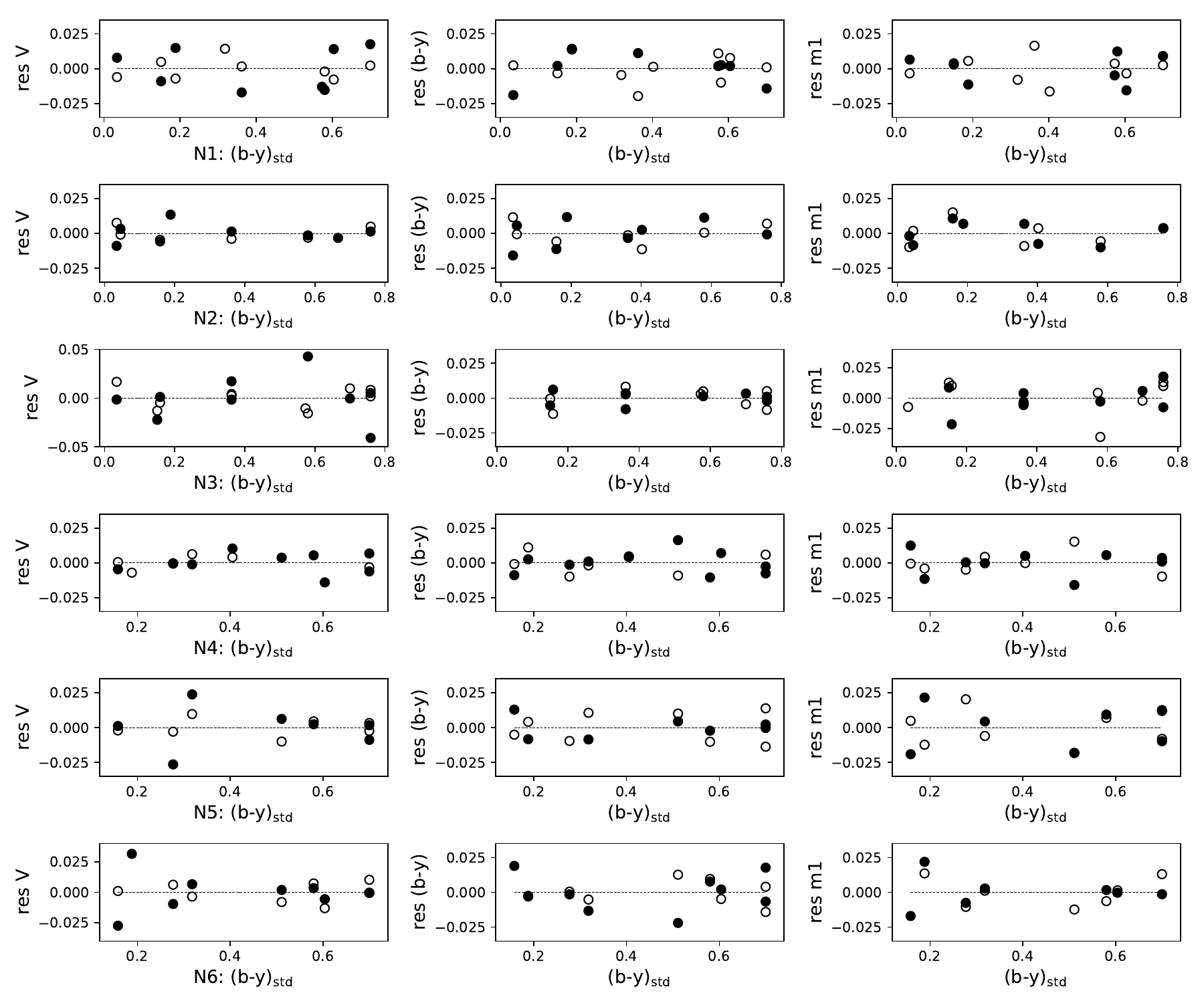}
      \caption{Residuals of the transformation to the standard system.
        Filled circles~are for chip~1 and open circles~for chip~2.
              }
      \label{fig:stdres}
   \end{figure}

   The magnitudes and colors of stars were standardized for each chip of the camera
   separately using the following transformation equations:

   \begin{align*}
     y_{\mathrm{inst}} &=  V_{\mathrm{std}} + a_1 + a_2 \cdot (b-y)_{\mathrm{std}} + a_3 \cdot (X_y - 1.25), \\
     (b-y)_{\mathrm{inst}} &= b_1 + b_2 \cdot (b-y)_{\mathrm{std}} + b_3 \cdot (X_{(b-y)} - 1.25), \\
     m1_{\mathrm{inst}} &= c_1 + c_2 \cdot (b-y)_{\mathrm{std}} + c_3 \cdot (X_{m1} - 1.25)
     + c_4 \cdot m1_{\mathrm{std}},
   \end{align*}

   \noindent where $y_{\mathrm{inst}}$, $(b-y)_{\mathrm{inst}}$, and $m1_{\mathrm{inst}}$ are
   instrumental magnitude and colors; $V_{\mathrm{std}}$, $(b-y)_{\mathrm{std}}$
   and $m1_{\mathrm{std}}$ are standard magnitude and colors from the \citet{Paunzen2015}
   catalog; $X$ is airmass; and $a_i$, $b_i$, $c_i$ are transformation coefficients
   summarized in Table~\ref{tab:std}. The typical error of the transformation is
   lower than 0.02~mag and is given in  Col. 8 of Table~\ref{tab:std}, also
   illustrated as the spread of points in Fig.~\ref{fig:stdres}.
   The astrometric solutions for the images in the $y$ filter were obtained based
   on Gaia DR2 catalog \citep{Gaia1,Gaia2} with subarcsec accuracy.

   We tested the completeness of our photometry by performing artificial star tests.
   We used the ADDSTAR routine of the DAOPHOT package to add randomly generated artificial
   stars to each image in $y$ filter. Their number was about 5\% of the
   number of stars found in a~given frame. We created 20 such images for each subframe
   of each field and performed the profile photometry on them. Next, we added the
   same list of stars to images in $b$ and $v$ filters and repeated the procedure.
   We calculated the retrieval rate of artificial stars. The results in every cluster
   are roughly similar. The retrieval rate for stars brighter than $V=13$~mag is
   about 86\% in the $y$ filter, 93\% in $b$, and almost 100\% in $v$. The lower
   number of retrieved stars in the $y$ filter might be due to the overexposition
   of stars. Finally, for stars between 13 and 19~mag, the range for which we perform
   metallicity calculations, completeness is about 100\% in all filters and then starts
   to drop, reaching practically zero for stars fainter than $V=22$~mag.

\subsection{Selection of cluster members and field stars}
\label{ssec:sel}

   An efficient way to separate cluster members from non-members is via  the
   proper motions (PMs) of stars. We cross-matched our master lists of stars with
   the Gaia DR2 catalog. Unfortunately, the accuracy of Gaia PMs for stars from
   our fields turned out to be insufficient for reliable membership determination.
   The average PM error is about 1~mas$\cdot$yr$^{-1}$ in RA and
   0.93~mas$\cdot$yr$^{-1}$ in DEC, which correspond to about 296~km$\cdot$s$^{-1}$
   and 275~km$\cdot$s$^{-1}$, if distance to SMC of 62.44~kpc is used \citep{Graczyk2020}.
   Nevertheless, we used Gaia data to reject galactic foreground stars having significant
   values of PMs. To determine obvious SMC non-members we used a~similar approach
   to that described in  \citet{Narloch2017}, among others, where   stars are rejected
   based on their location
   on the vector point diagram. For all stars in a~given cluster field, we calculated
   mean values and standard deviations of their PMs,  $\mu$ (M$\alpha$, M$\delta$,
   S$\alpha$, S$\delta$), and PM errors,~$\sigma_{\mu}$ (ME$\alpha$, ME$\delta$,
   SE$\alpha$, SE$\delta$). We did not divide stars into magnitude bins because
   often there were  too few  stars in a~given bin. Next, we selected only stars satisfying
   the conditions $\mu \leq 3 \cdot S$ and $\sigma_{\mu} \leq ME + 3 \cdot SE$ and
   repeated the procedure. This way galactic foreground stars with high PMs were
   removed from the input lists.

   In the next step, we selected stars enclosed in a certain radius as cluster members.
   We adopted equatorial coordinates and radii of clusters from updated catalog of
   \citet{Bica2020}. Stars outside this cluster radius were classified as field stars.

   We decided to not perform a~statistical subtraction of the field stars.
   The number of stars in most clusters is small and they are located in dense fields
   which makes it difficult to do statistical subtraction correctly. The
   small field of view of the camera does not  provide good statistics for
   the field stars, and  we cannot be sure that there are  no cluster members among them.
   On the other hand, in star clusters located farther from the SMC center in sparse
   fields, the contamination of the field stars is negligible. Once the
   individual metallicities are derived, they can help to disentangle field and
   cluster members.

\subsection{Determination of the reddening toward clusters}
\label{ssec:red}

   In order to correct data for the reddening we used reddening maps published recently
   by \citet[][hereafter G20]{Gorski2020}, and \citet[][hereafter S21]{Skowron2020},
   both obtained by calculating the difference of the observed and intrinsic color
   of the RC stars in the SMC; G20 used the OGLE-III data set while S21 used OGLE-IV
   data with a~much larger field of view. The reddening values for our fields from
   S21 are systematically smaller than those from G20 (see Table~\ref{tab:lit}).
   We adopted the average of both maps ($E(B-V)_{GS}$) as the reddening
   of the stellar clusters and their surrounding fields. The adopted reddenings
   are independent of our data. The $E(V-I)$ from the S21 maps were converted into
   $E(B-V)$ with $E(B-V) = E(V-I)/1.318$. In the cases of Lindsay~1, Lindsay~113,
   and NGC~339, which were out of reach of G20 maps, we only applied the reddening
   from S21.
   Typical uncertainties of the reddening values in a~given field are dominated
   by systematic errors of the apparent color measurements, equal to about 0.013~mag
   in G20. In S21 the error on the intrinsic color in the SMC is about 0.016~mag.
   The average difference between G20 and S21 reddening values for fields studied
   in this work is about 0.031~mag. Half of this value, rounded up, yielded
   $\sigma_{E(B-V)_{GS}} = 0.016$~mag, which was propagated into the systematic
   error on the derived metallicities resulting from the reddening.

   We calculated the reddening values for magnitudes and colors using the following equations:
   $A_V = 3.315 \cdot E(B-V)$, $E(b-y) = 0.772 \cdot E(B-V),$ and
   $E(m1) = -0.269 \cdot E(B-V)$ \citep{Schlegel1998}.

\subsection{Metallicity calculation based on Str\"omgren colors}
\label{ssec:metal}

   A~convenient property of the Str\"omgren photometric system, which makes it
   very useful in stellar astrophysics, is the possibility to obtain the metallicity
   for individual stars nearly independent of their age  \citep[e.g.,][]{Dirsch2000}.
   For the calculation of the metallicity we adopted a~calibration of the Str\"omgren
   $m1-(b-y)$ two-color relation derived by \citet{Hilker2000}. This relation is
   valid only in the certain color range  $0.5<(b-y)<1.1$. The calibration equation
   is

   \begin{equation}
   \label{eq:feh}
     \mathrm{[Fe/H]} = \frac{m1_0 + a1 \cdot (b-y)_0 + a2}{a3 \cdot (b-y)_0 + a4}
   ,\end{equation}

  \noindent where

   $$a1 = -1.277 \pm 0.050,\, a2 = 0.331 \pm 0.035,$$
   $$a3 = 0.324 \pm 0.035,\, a4 = -0.032 \pm 0.025.$$

   Errors on the metallicity determination of individual stars can be calculated by
   performing full error propagation of Equation~\ref{eq:feh} as  \citep{Piatti2019}

   \begin{align}
   \label{eq:feher}
     \sigma_{\mathrm{[Fe/H]}} &=  \biggl[ \left(\frac{(b-y)_0}{c}\sigma_{a1}\right)^2
                     + \left(\frac{1}{c}\sigma_{a2}\right)^2 \notag
                     + \left( \frac{(b-y)_0 \cdot \mathrm{[Fe/H]}}{c}\sigma_{a3}\right)^2 \\
                     &+ \left( \frac{\mathrm{[Fe/H]}}{c}\sigma_{a4}\right)^2
                     + \left( \frac{(a1 - a3 \cdot \mathrm{[Fe/H]})}{c}\sigma_{(b-y)_0}\right)^2 \\
                     &+ \left( \frac{1}{c}\sigma_{m1_0}\right)^2 \biggr]^{\frac{1}{2}}, \notag
   \end{align}

   where

   $$c = a3 \cdot (b-y)_0 + a4. $$

   The first four terms in Equation~\ref{eq:feher} relate to the systematic error,
   and the remaining two to the statistical error.
   The dominant error in the Str\"omgren two-color diagram is $\sigma_{m1_0}$.
   Consequently, the metallicity error is larger for more metal-poor stars than
   for the more metal-rich correspondents of the same color.
   The calibration of Hilker (2000) was based on stars, with spectroscopic
   metallicities on the \citet[][hereafter the ZW84 scale]{ZW1984}.

   In the first step of the metallicity determination we selected stars from the dereddened
   (see Section~\ref{ssec:red}) color range of $0.5<(b-y)_0<1.1$. Moreover, only
   stars having $\sigma_{(b-y)_0}<0.1$ and $\sigma_{m1_0}<0.1$ (as calculated by
   DAOPHOT) were accepted for further calculations.
   Even so, on the $m1_0 - (b-y)_0$ relation some of the  bluest stars from this color range
   deviate greatly from the well-defined relation for redder stars. In addition, the applied ~cut
   in color on the blue edge introduces a~bias toward metal-poor stars with larger
   metallicity errors. To work around this problem, and to eliminate deviating stars,
   we followed \citet{Dirsch2000} and applied additional selection criteria by drawing
   a~line perpendicular to the base of the RGB and included only stars
   redder than this line. In the first iteration the mean and unbiased standard
   deviation for the remaining stars were calculated. Next, we applied 3$\sigma$ clipping
   to reject outliers and recalculated the previously obtained values. In the end,
   the resulting CMDs and $m1_0 - (b-y)_0$ relations were examined by eye and single
   stars deviating significantly from either of them were rejected manually and the final
   values of the mean and the unbiased standard deviation were obtained. As a~statistical
   error of the mean metallicity we adopted unbiased standard deviation divided
   by the square root of the number of stars used for the calculation.

\begin{figure}
   \centering
   \includegraphics[width=\hsize]{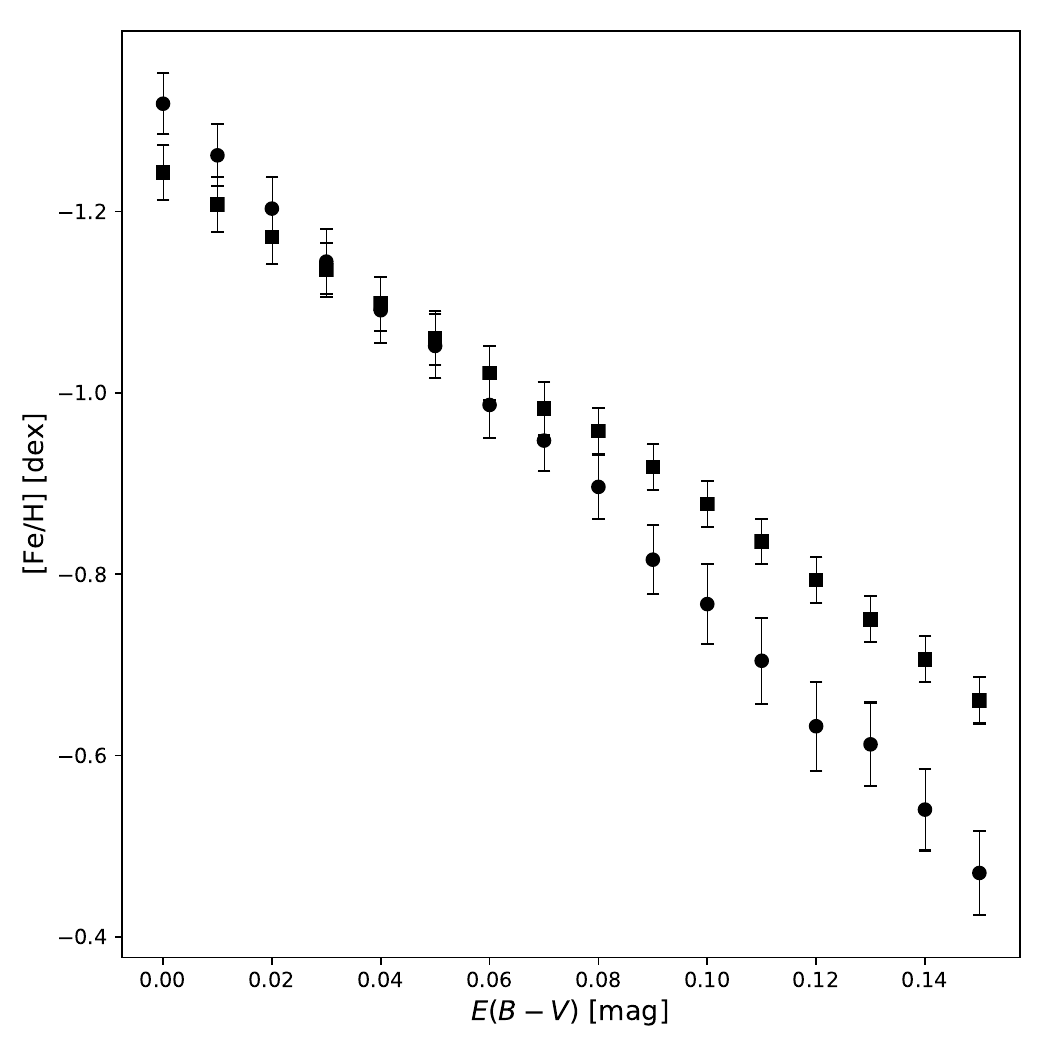}
      \caption{Metallicity derived from Eq. (\ref{eq:feh}) vs. assumed
      reddening in Lindsay~113 (points) and NGC~330 (squares).
              }
      \label{fig:fehredd}
   \end{figure}

   \citet{Dirsch2000} noted that metallicities measured photometrically are very
   sensitive to assumed reddening, being the major source of systematic metallicity
   error.
   This degeneracy is particularly important in old clusters and field stars. In younger
   clusters the reddening can be determined quite precisely because  the color of the hot main
   sequence stars depends weakly on the temperature and metallicity.
   Figure~\ref{fig:fehredd} shows how [Fe/H] depends on the $E(B-V)$ for two star
   clusters: the older Lindsay~113 and the younger the NGC~330. The errors shown in
   the figure were derived by dividing the unbiased standard deviation of the mean
   metallicity of a~cluster by the square root of the number of stars used for the
   metallicity calculation.
   The increase in  assumed reddening of $0.01$~mag increases the derived metallicity
   by about 0.06~dex for Lindsay~113 and 0.04~dex for NGC~330, so on average
   by about 0.05~dex. For a~typical error on the adopted reddening (see
   Section~\ref{ssec:red}), this corresponds to $\sigma_{\mathrm{[Fe/H]}} \approx
   0.08$~dex for all clusters and surrounding fields.

   The second source of systematic uncertainty is the precision of the $m1$ and $(b-y)$
   calibration to the standard system. The calibration errors cause a bias of the
   corresponding metallicity that depends on  the color of a~star. This effect is
   more profound for bluer stars, which leads to larger metallicity errors
   \citep[see Fig.~1 in][]{Dirsch2000}.
   As described in Sect.~\ref{sec:obsred}, the r.m.s. errors for the calibration of
   $(b-y)$ and $m1$ for all nights are smaller than 0.02~mag, so we assume that
   this is a~maximum uncertainty for these values. The metallicity error for a~given
   cluster or field arising from transformation to the standard system can be derived
   from simulations. On each night we draw a~random number from the normal
   distribution with standard deviation equal to the r.m.s of a~given chip (from
   Table~\ref{tab:std}) and add it to the $(b-y)$ colors of stars used for the mean
   metallicity calculation of a~given cluster or field. We do the same for $m1$
   index. We then calculate the new mean metallicity. We repeat this procedure
   10000 times and determine the mean and standard deviation of obtained
   new mean metallicities from the simulations.
   We then adopt this standard deviation as the error resulting from the
   transformation to the standard system for a~given stellar cluster or field.

   Another source of uncertainty is differential reddening across the field, which
   might affect stars with different colors in a~different way, resulting in broadening
   of the metallicity distribution and consequently also the age distribution.
   Due to the small field of view of the SOAR telescope and the resolution of the G20
   reddening maps of 3~arcmin, and of the S21 maps of about 1.7~arcmin in the central parts
   of the SMC, decreasing in the outskirts, we cannot precisely estimate this effect.
   Consequently, we neglected it.

   Finally, the total systematic error of a~given mean metallicity is composed of
   reddening and calibration errors added in squares under the square root. The
   statistical and systematic errors for the cluster and field samples
   are given in Tables~\ref{tab:reddC} and \ref{tab:reddF}, respectively.

%

\subsection{Age determination}
\label{ssec:age}

   For the age determination we employed isochrones from the Dartmouth Stellar
   Evolutionary Database\footnote{http://stellar.dartmouth.edu/models/isolf\_new.html}
   \citep[][hereafter the Dartmouth isochrones]{Dotter2008} and the Padova database of
   stellar evolutionary tracks and isochrones available through the CMD 3.3
   interface\footnote{http://stev.oapd.inaf.it/cgi-bin/cmd\_3.3} \citep{Marigo2017}
   calculated with the PARSEC \citep{Bressan2020} and COLIBRI \citep{Pastorelli2019}
   evolutionary tracks (hereafter the Padova isochrones).
   The Dartmouth isochrones were available only for the $1-15$~Gyr isochrones sets,
   so were too old for most of our clusters. Instead, the Padova isochrones
   covered all possible ages so most of our results are based on these isochrones.
   Still, in cases where it was possible, we used both types of isochrones for
   the comparison. The Dartmouth isochrones seem to better reflect the shape of red
   giant branches in older clusters, while  the Padova isochrones tend to flatten too
   much near the tip.

   During the isochrone fitting procedure, we   used the isochrone of a~specific
   metallicity determined from the Str\"omgren data at a~fixed distance to the SMC
   from \citet{Graczyk2020}, where the distance modulus is $(m-M)_{\mathrm{SMC}} =
   18.977$~mag.
   In a few cases (Lindsay~6, Lindsay~113, NGC361, IC1611) the isochrones of the adopted
   metallicity did not fit well to the CMD of the cluster, so we adopted new values
   for the reddening from the literature, recalculated the metallicity, and repeated
   the procedure.
   The error in age was adopted as a~half of the age difference between two marginally
   fitting isochrones selected around the best fitting isochrone.

\subsection{Str\"omgren photometry}
\label{ssec:phot}

   Table~\ref{tab:phot} presents the first five rows of the compiled catalog of Str\"omgren
   photometry for stars from the fields studied in this work. The catalog contains
   stars for which all three Str\"omgren $vby$ filters were available, and subsequently
   the color indices $(b-y)$ and $m1$ could be calculated. The photometric errors presented
   in the table   result from the DAOPHOT package and the full error propagation
   of the transformation equations with coefficients from Table~\ref{tab:std}, given
   in a~separate column.
   A~complete version of Table~\ref{tab:phot} is available online on the
   Araucaria Project webpage\footnote{https://araucaria.camk.edu.pl/} and at the
   CDS via anonymous ftp to cdsarc.u-strasbg.fr (ftp://130.79.128.5) or via
   http://cdsarc.u-strasbg.fr/viz-bin/cat/J/A+A/647/A135

\section{Results}
\label{sec:results}

\begin{figure*}
   \resizebox{\hsize}{!}
            {\includegraphics[]{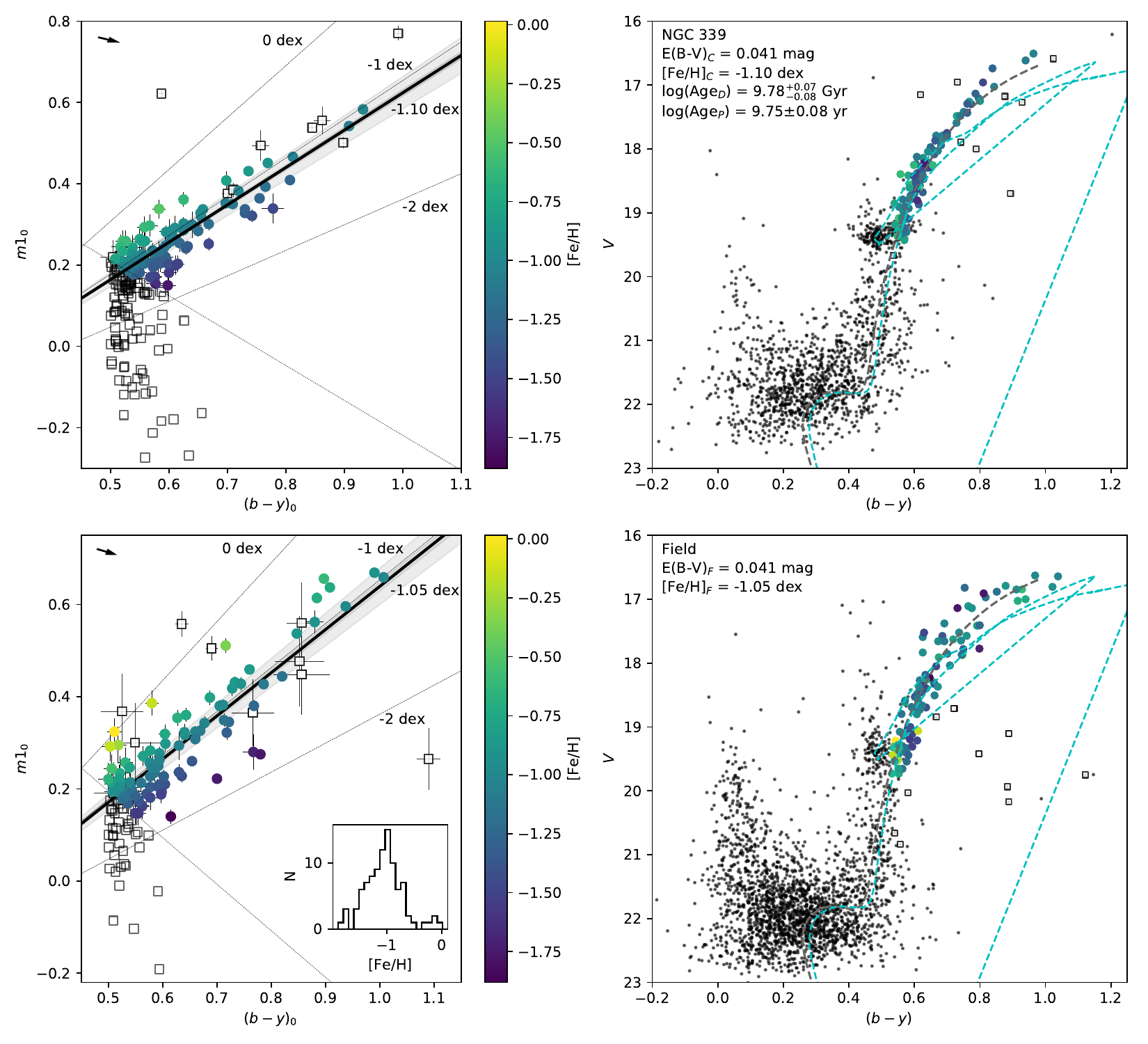}}
      \caption{Reddening corrected two-color diagrams and reddened CMDs for
               NGC~339 (upper panels) and its surrounding field stars (lower panels).
               Left panels:
                Stars having photometry in $vby$ filters (black points);
               stars excluded from metallicity determination (open squares);
               stars used to calculate the mean metallicity of a~cluster
               and field  (color-coded points, where colors represent metallicity distribution);
               lines of constant metallicity (dashed lines);
               additional selection criteria drawn after visual inspection
               of the plot (dotted line);
               obtained mean metallicities of cluster and field stars (black solid lines);
               the statistical and systematic errors of
               the mean metallicity of the cluster (darker and lighter shaded areas).
               The arrows indicate the reddening vectors.
               Right panels: Dartmouth and Padova
               best-fitting isochrones (gray and turquoise dashed lines, respectively) superimposed
               onto the field CMD,
               in order to illustrate the position of the cluster against field stars.
              }
      \label{fig:ngc339}
   \end{figure*}

\begin{figure*}
   \resizebox{\hsize}{!}
            {\includegraphics[]{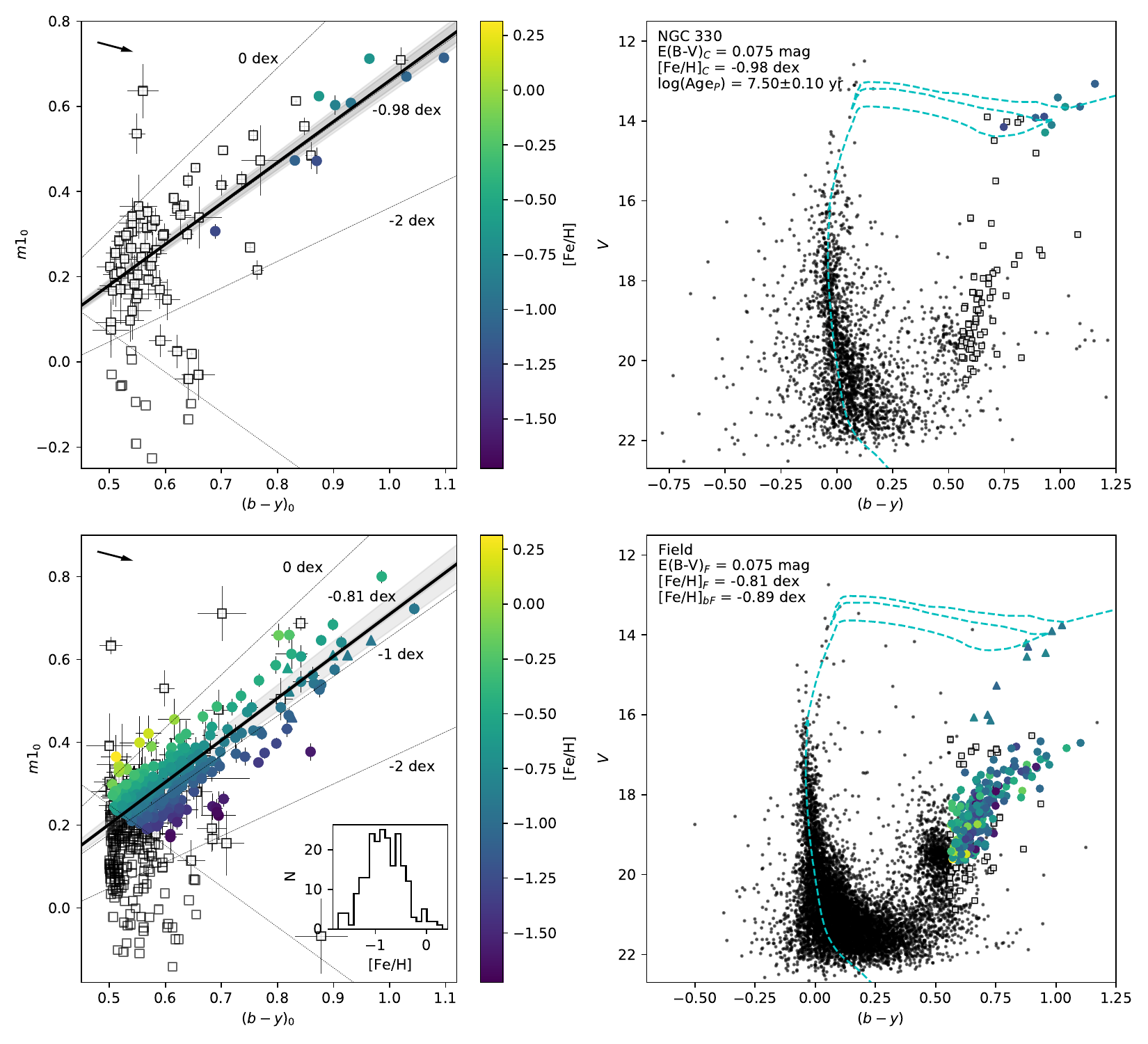}}
      \caption{Reddening corrected two-color diagrams and reddened CMDs for
               NGC~330 (upper panels) and surrounding field stars (lower panels).
               Triangles indicate the young field giants. The meaning of
               other symbols is the same as in Fig.~\ref{fig:ngc339}.
              }
      \label{fig:ngc330}
   \end{figure*}

\begin{figure*}
   \resizebox{\hsize}{!}
            {\includegraphics[]{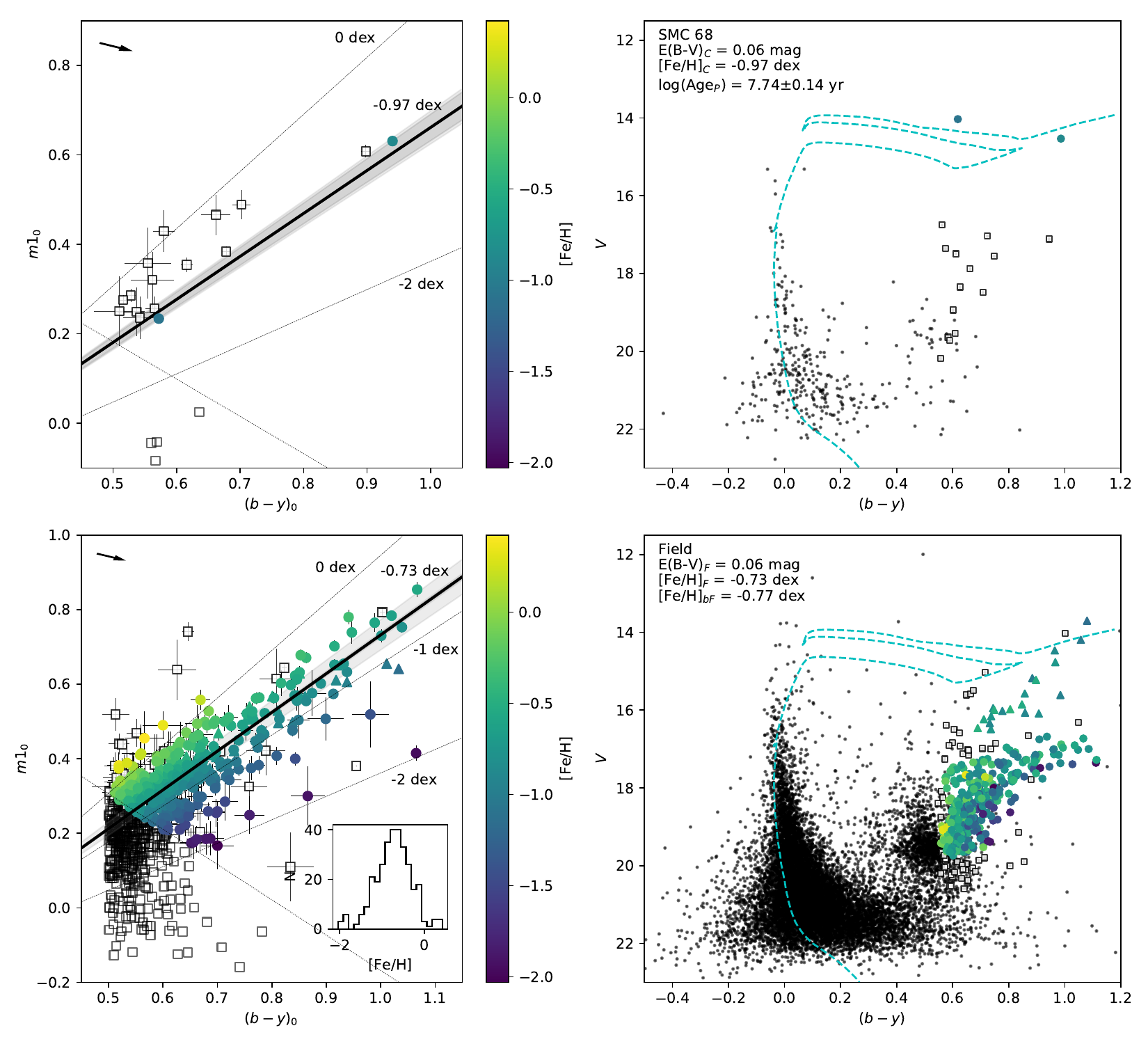}}
      \caption{Reddening corrected two-color diagrams and reddened CMDs for
               OGLE-CL~SMC~68 (upper panels) and surrounding field stars (lower panels).
               Triangles indicate the young field giants. The meaning of
               other symbols is the same as in Fig.~\ref{fig:ngc339}.
              }
      \label{fig:smc68}
   \end{figure*}

\begin{figure*}
   \resizebox{\hsize}{!}
            {\includegraphics[]{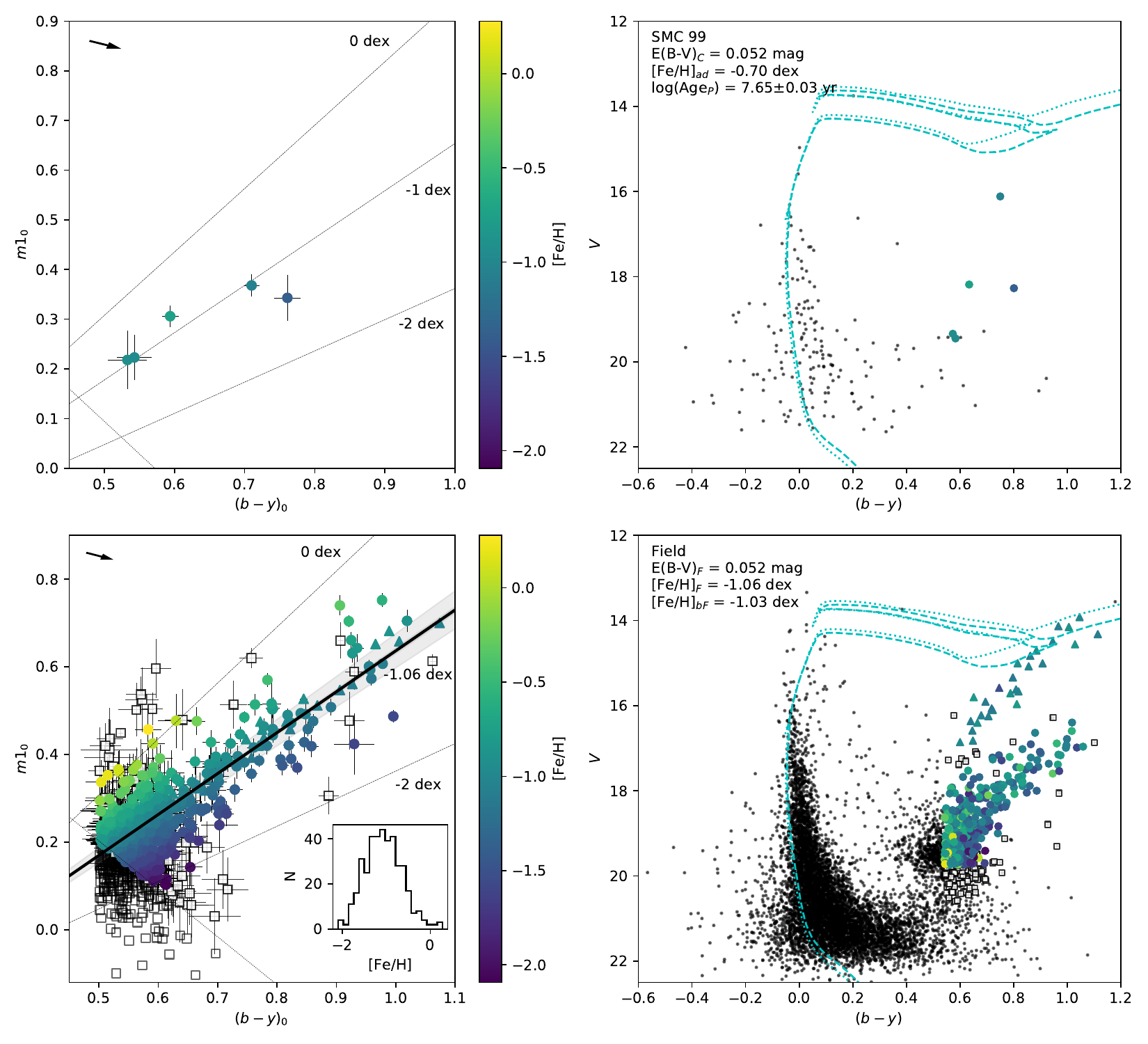}}
      \caption{Reddening corrected two-color diagrams and reddened CMDs for
               OGLE-CL~SMC~99 (upper panels) and surrounding field stars (lower panels).
               Triangles indicate the young field giants. The meaning of
               other symbols is the same as in Fig.~\ref{fig:ngc339}.
               The turquoise dashed line is a~Padova best-fitting isochrone for the adopted
               metallicity and metallicity of young field giants.
              }
      \label{fig:smc99}
   \end{figure*}

\begin{figure}
   \centering
   \includegraphics[width=\hsize]{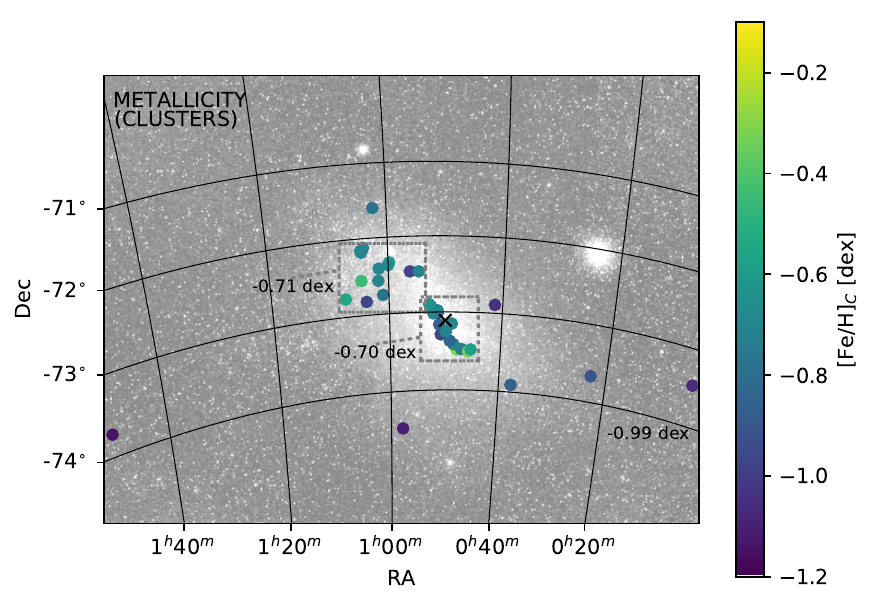}
   \includegraphics[width=\hsize]{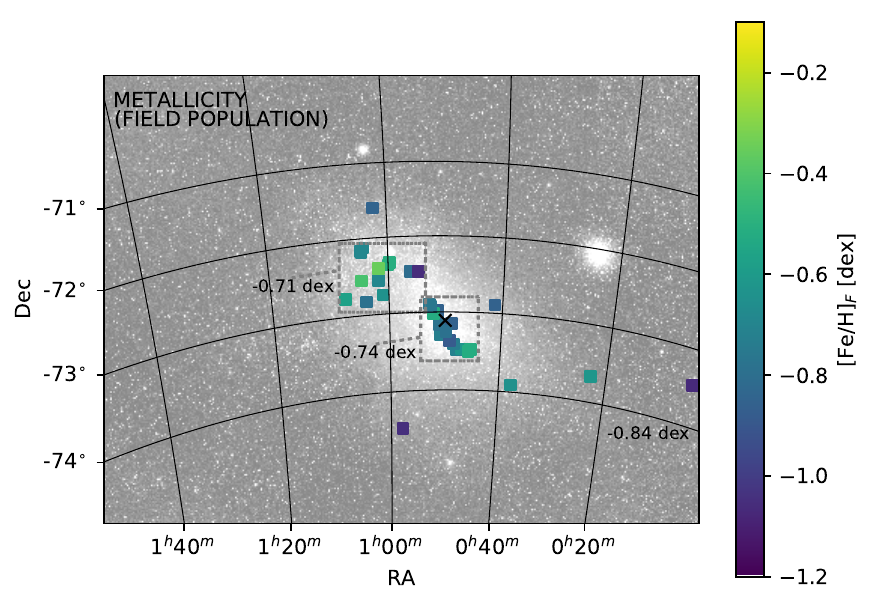}
      \caption{Metallicity map of the field (upper panel) and cluster (lower panel)
               stars in the SMC. The black cross indicates the center of the SMC at
               $\alpha_0 = 12.54$~deg; $\delta_0 = -73.11$ \citep{Ripepi2017}.
               North is up; east is left.
              }
      \label{fig:mapfeh}
   \end{figure}

\begin{figure}
   \centering
   \includegraphics[width=\hsize]{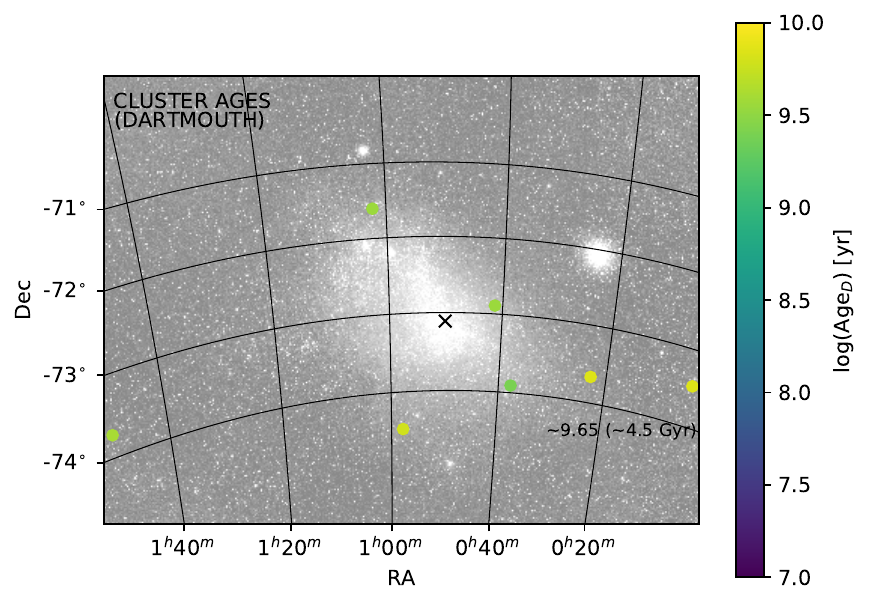}
   \includegraphics[width=\hsize]{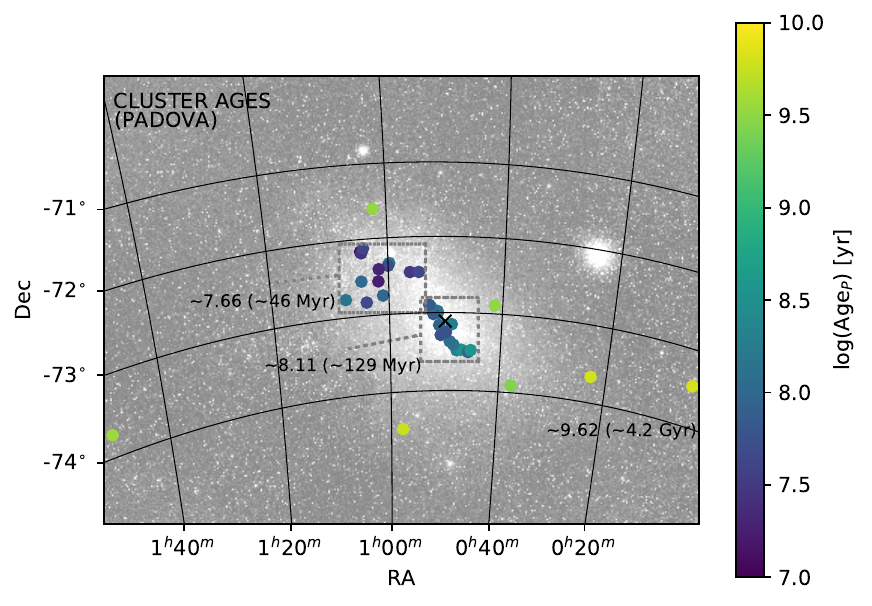}
      \caption{Age map of the cluster stars in the SMC derived from the Dartmouth
               (upper panel) and Padova (lower panel) theoretical isochrones.
               The black cross indicates the center of the SMC as in Fig.~\ref{fig:mapfeh}.
               North is up; east is left.
              }
      \label{fig:mapage}
   \end{figure}

   Figures~\ref{fig:ngc339}--\ref{fig:smc99} present two-color diagrams and CMDs
   for stellar clusters and their surrounding fields for three cases: clusters with
   well-populated RGBs (Fig.~\ref{fig:ngc339}), and  young clusters with a~few stars
   (Fig.~\ref{fig:ngc330}--\ref{fig:smc68}) and no stars (Fig.~\ref{fig:smc99})
   for metallicity calculation.
   Figure~\ref{fig:mapfeh} shows the spatial distribution of the mean metallicities
   of clusters and fields, while Fig.~\ref{fig:mapage} similarly shows the
   distribution of cluster ages. We summarized our measurements in Tables~\ref{tab:reddC}
   (clusters) and \ref{tab:reddF} (fields).

\begin{figure}
   \centering
   \includegraphics[width=\hsize]{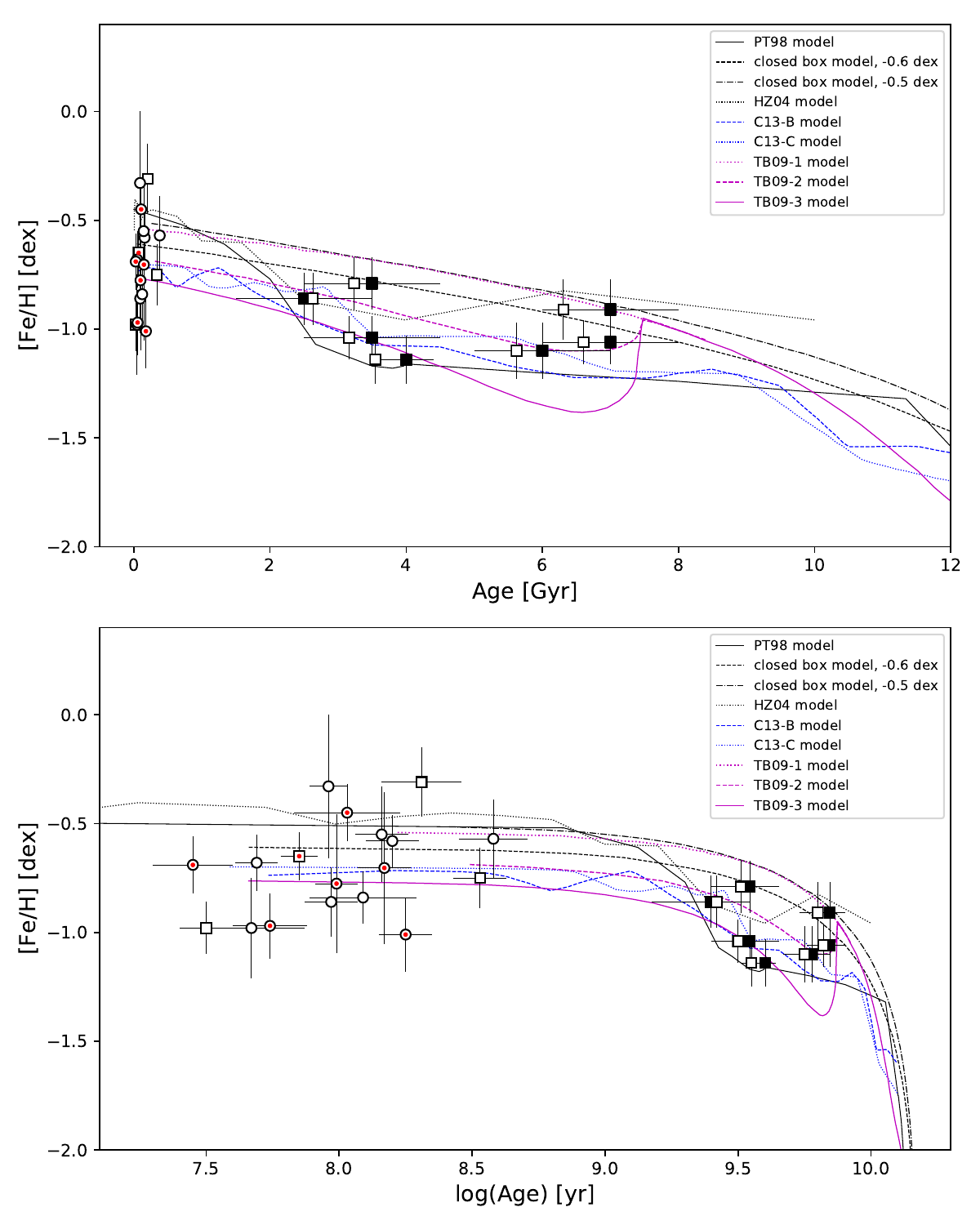}
      \caption{Age--metallicity relation for clusters studied in this work.
               Clusters with ages derived using Dartmouth isochrones (black squares);
               clusters with reliable number of stars for metallicity determination
               (open squares), having ages derived from the Padova isochrones;
               clusters with 1--4 stars for metallicity calculation
               with the Padova ages (open circles). Red dots indicate clusters
               whose metallicity was determined for the first time.
               Overplotted theoretical models:  PT98 bursting model (solid line);
               closed box model for $-0.6$~dex and $-0.5$~dex (dashed and dash-dotted
               line, respectively); HZ04 model (dotted line); C13-B and C13-C models
               (blue dashed and dotted lines, respectively);  TB09-1, TB09-2, and
               TB09-3 models (magenta dotted, dashed, and solid lines, respectively).
              }
      \label{fig:am}
   \end{figure}

   There are seven intermediate-age stellar clusters in our sample of 35 SMC clusters
   (Lindsay~1, 6, 19, 27, and 113; NGC~339 and 361) with well-populated RGBs having
   between 18 and 93 stars used for metallicity determination. These are well studied
   objects having many age and metallicity measurements in the literature.
   For their age determination the Dartmouth and the Padova isochrones were both used.
   A~further six clusters in our sample (NGC~330 and 265; OGLE-CL~SMC~45, 69, and
   88; IC~1611) are young stellar clusters with between 4 and 12 red giants, which
   is sufficient for reliable metallicity determination. Only the  Padova isochrones
   were used for their age determination. We indicate these 13 clusters in
   Fig.~\ref{fig:am} with squares.
   Twelve of the young star clusters studied in this work had less than four, but there was at
   least one star lying within the cluster radius and fulfilling the criteria for
   metallicity determination. To the best of our knowledge, the metallicities for
   OGLE-CL~SMC~68, 71, 82, 126, 143, and [BS95]~123 are provided for the first time.
   NGC~376, IC~1612, Bruck~39, and OGLE-CL~SMC~32, 54, and 156 have at least one previous
   metallicity estimation. Only the Padova isochrones were suitable for their age determination.
   Clusters from this group are indicated in Fig.~\ref{fig:am} with open circles.
   Finally, there were ten young star clusters (OGLE-CL~SMC~49, 50, 61, 78, 99, 128,
   129, 142, 144, and 205) for which we have not found any suitable stars for cluster
   metallicity calculation. Moreover, these clusters have no published metallicities
   in the literature,  the sole exception being OGLE CL-SMC-49 with one such
   determination. However, we were able to calculate the mean metallicities of the field
   stars surrounding these clusters. We also estimated the ages of the clusters by
   employing the Padova isochrones for $\mathrm{[Fe/H]} = -0.70$~dex, which is a~value
   close to the mean metallicity of the SMC (discussed in the next section).

%

%

%

%

   \begin{figure}
   \centering
   \includegraphics[width=\hsize]{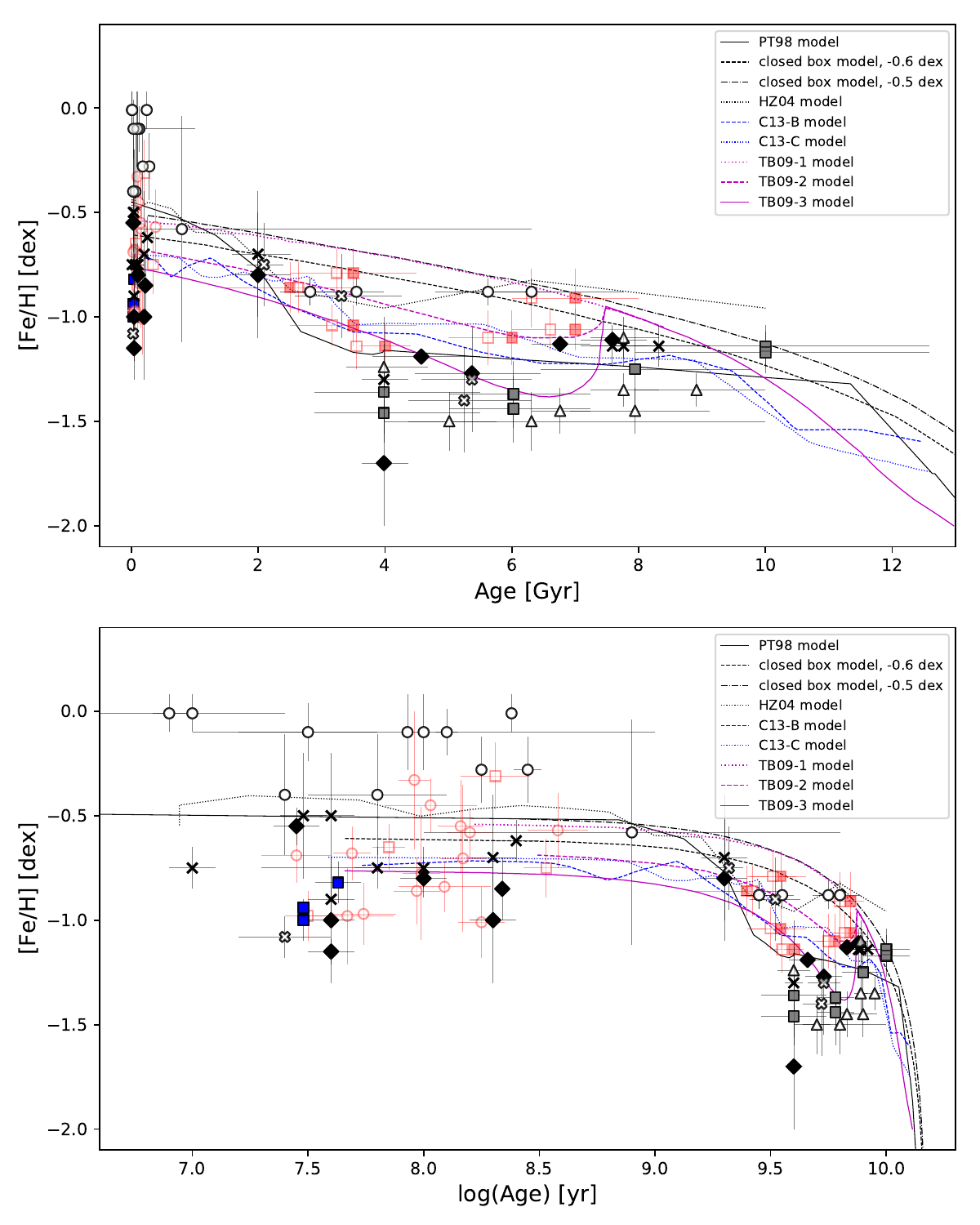}
      \caption{Age--metallicity relation for clusters studied in this work compared
               with the literature (see Table~\ref{tab:lit}).
               Values derived from two-color Str\"omgren diagram (black diamonds);
               values derived by \citet{Perren2017} using the ASteCA package (open circles);
               low-resolution spectroscopic metallicities expressed in ZW84 scale (gray squares);
               high-resolution spectroscopic metallicities in ZW84 scale (blue squares);
               values obtained from RGB slope method given in ZW84 scale (open triangles);
               values derived from fitting of theoretical isochrones to optical data (crosses);
               values derived from fitting of theoretical isochrones to data
               in Washington system (open crosses).
               Red squares and open circles indicate measurements from this work presented in
               Fig.~\ref{fig:am} for comparison.
               Overplotted theoretical models are as in Fig.~\ref{fig:am}.
              }
      \label{fig:amlit}
   \end{figure}

   \begin{figure}
   \centering
   \includegraphics[width=\hsize]{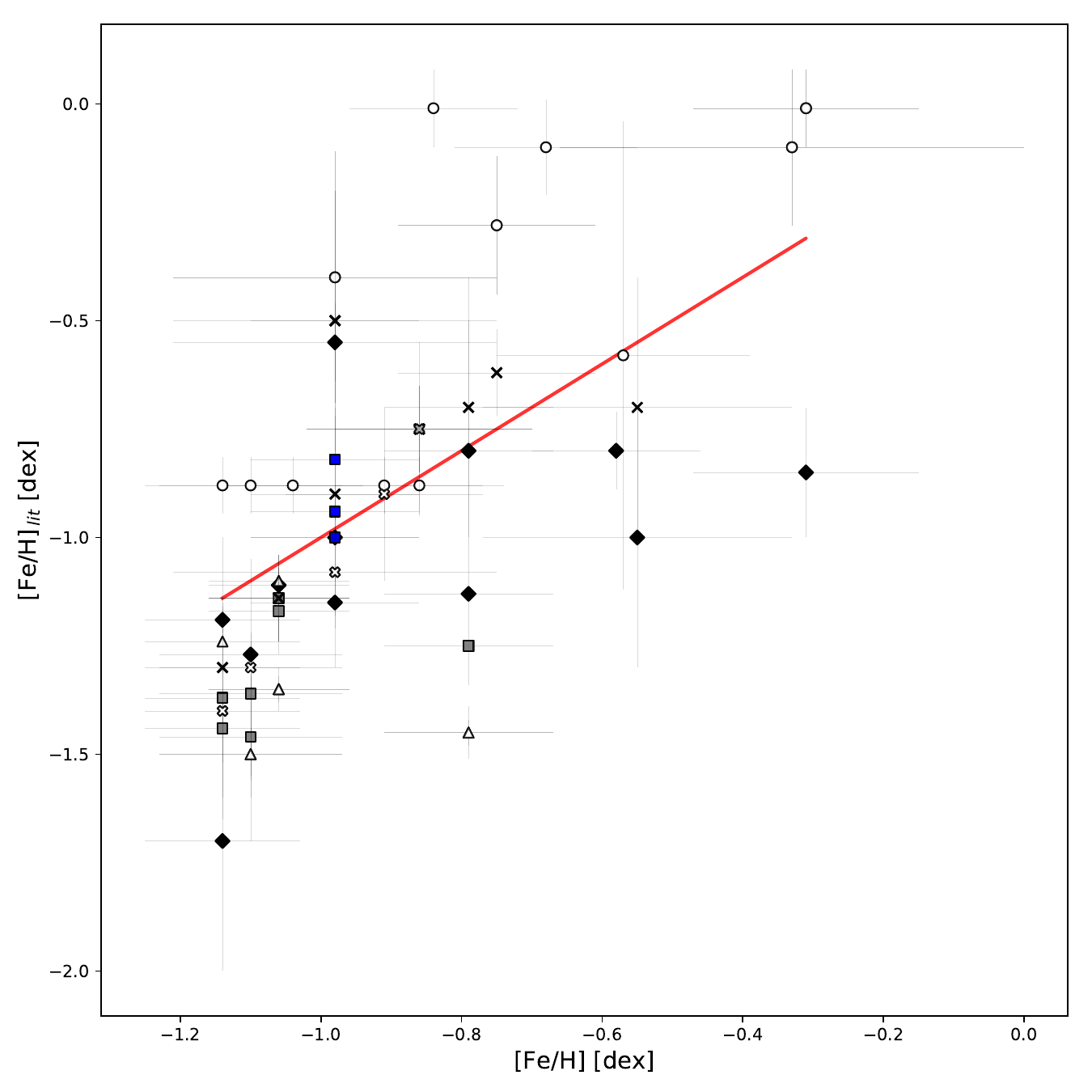}
      \caption{Comparison of the metallicities for star clusters obtained in
      this work with the literature values from Fig.~\ref{fig:amlit}. The red solid
      line represents the 1:1 relation.
      }
      \label{fig:fehlit}
   \end{figure}

\subsection{Metallicity distribution for cluster and field stars in the SMC}
\label{ssec:metalmap}

   The upper panel in Fig.~\ref{fig:mapfeh} presents a~metallicity map of the
   star clusters given in Table~\ref{tab:reddC}. The black cross indicates the center
   of the SMC determined based on the distribution of classical Cepheid variables
   by \citet{Ripepi2017} ($\alpha_0 = 12.54 \pm 0.01$~deg; $\delta_0 = -73.11 \pm
   0.01$~deg). In the map it is clear that the most metal-poor clusters are located
   in the outskirts of the galaxy, while the more metal-rich ones group close to
   the SMC central region.
   The average metallicity of the clusters most distant from the SMC center (Lindsay~1,
   6, and 113, and NGC~336) is $-1.05$~dex with a~standard deviation ($\sigma$) of
   0.10~dex. The average metallicity of three clusters located a~bit farther from
   the elongated central region (NGC~361, Lindsay~19 and 27) is $-0.90$~dex ($\sigma
   = 0.13$~dex). The average metallicity of all seven outer clusters is $-0.99$~dex
   ($\sigma = 0.13$~dex). Most of the clusters studied in this work are located
   along the denser central region of the SMC. Their average metallicity is
   $-0.70$~dex ($\sigma = 0.22$~dex).
   Moreover, we can distinguish two groups, one positioned close to the SMC center
   to the west (hereafter western group) and a~more sparse group located
   where the HI super-shell 304~A is placed (hereafter eastern group).
   We show these two groups in Fig.~\ref{fig:mapfeh} with dotted rectangles.
   Five clusters from the western group (OGLE-CL~SMC~49, 50, 61, 78, and
   205) had no suitable stars for the metallicity calculation, so no information
   about their metallicity is available. The average metallicity of ten remaining
   clusters NGC~265; OGLE-CL~SMC~32, 45, 54, 68, 69, 71, 82, 88; and Bruck~39) is
   $-0.70$~dex ($\sigma = 0.24$~dex). In case of eight clusters from the
   eastern group (NGC~330 and 376; IC~1611; OGLE-CL~SMC~126, 143, and 156)
   the average metallicity is $-0.71$~dex ($\sigma = 0.19$~dex) where clusters
   OGLE-CL~SMC~99, 128, 129, 143, and 144 were omitted because we had no information
   about their metallicities. The values for the eastern and western groups
   agree within the errors.

   The lower panel in Fig.~\ref{fig:mapfeh} presents a~map of the mean metallicities
   of the fields surrounding the studied star clusters given in Table~\ref{tab:reddF}.
   In five cases we studied more than one cluster in the field of view, so the
   field stars from these regions come from the area around all present clusters.
   The average metallicity of the old giants used for the metallicity calculation
   from the surrounding of the three  outermost clusters (without Lindsay~113, where
   there are only a~few stars most probably belonging to the cluster) is $-0.93$~dex
   ($\sigma = 0.39$~dex). For field stars around the three other clusters located
   closer to the elongated SMC central region it is $-0.80$~dex ($\sigma = 0.36$~dex).
   The average metallicity of old giants surrounding all six intermediate-age clusters
   is $-0.84$~dex ($\sigma = 0.37$~dex). The average metallicity of the field stars
   used for the metallicity calculation lying in the central regions of the SMC
   is $-0.73$~dex ($\sigma = 0.45$~dex).
   For the western group it is $-0.74$~dex ($\sigma = 0.46$~dex) and for the eastern
   $-0.71$~dex ($\sigma = 0.43$~dex), which is a statistically insignificant difference.

   The young helium burning giants (HBGs) belonging to the field are common in the
   central regions of the SMC, but in the outer fields in our sample they are
   absent, or nearly so.
   In this population of stars we can expect to observe Cepheid variables.
   \citet{Ripepi2017} calculated photometric metallicities for 462 Cepheids and
   found a~peak of their metallicity distribution at about $\mathrm{[Fe/H]} = -0.60$~dex.
   On the other hand, \citet{Romaniello2008} reported $\mathrm{[Fe/H]} = -0.75 \pm
   0.02$~dex (with dispersion of 0.08~dex) based on high-resolution spectroscopic
   studies of 14 stars from the SMC.
   Recently, \citet{Lemasle2017} published metallicities for four SMC Cepheids
   obtained from high-resolution spectroscopy with average of $\mathrm{[Fe/H]} =
   -0.74$~dex. The average metallicity of all HBG stars measured by us in the fields
   around clusters is $-0.70$~dex ($\sigma = 0.28$~dex), which is in excellent agreement
   with the mentioned studies. The difference between the western ($-0.69$~dex,
   $\sigma = 0.26$~dex) and the eastern ($-0.71$~dex, $\sigma = 0.32$~dex) group
   is statistically insignificant.

\subsection{Age distribution for clusters in the SMC}
\label{ssec:agemap}

   Figure~\ref{fig:mapage} shows the age map of studied clusters in the SMC derived
   from the Dartmouth (upper panel) and Padova (lower panel) isochrones given in
   Table~\ref{tab:reddC}. The rectangles superimposed onto the plot indicate the same
   spatial groups of clusters discussed in previous subsection. The outermost clusters
   in our sample   are the oldest ones. The logarithm of their ages derived from the
   Dartmouth isochrones ranges between 9.40 to 9.85 (2.5 to 7.0~Gyr) with the average
   of 9.65 with $\sigma = 0.16$ ($\sim$4.5~Gyr) and from the Padova isochrones between
   9.42 to 9.82 (2.6 to 6.6~Gyr) with the average of 9.62 and $\sigma = 0.15$
   ($\sim$4.2~Gyr).
   For the central clusters only the Padova isochrones were available. The average logarithm
   of ages derived from the Padova isochrones for the western group of clusters
   is 8.11 with $\sigma = 0.26$ ($\sim$129~Myr) and for the eastern 7.66 with
   $\sigma = 0.36$ ($\sim$46~Myr). The mean age for both groups of central
   young clusters is 7.90 with $\sigma = 0.38$ ($\sim$79~Myr). The spatial age distribution
   we obtained closely  follows  the distribution of young star clusters presented by
   \citet[][see their Fig.~7]{Glatt2010} where most of the young clusters are
   located along the central overdensity and to the east from the center of the SMC
   where the HI super-shell 304~A is placed.

\subsection{Age--metallicity relation for clusters in the SMC}
\label{ssec:amrel}

   The resulting age--metallicity relation for star clusters studied in this work
   is illustrated in Fig.~\ref{fig:am}.

   In order to compare our results with theoretical predictions of the chemical
   evolution of the SMC, we considered a~few models published in the literature.
   The model of \citet[][hereafter the PT98 model,]{PG1998} predicts intensive star
   formation and chemical enrichment in the SMC during the initial epoch that
   brought the metallicity of the galaxy up to about $-1.3$~dex.
   This turbulent period was subsequently followed by relative quiescence lasting
   for about 8~Gyr, to finally be disturbed by rapid burst of chemical enrichment
   about 3~Gyr ago, which brought the global metallicity up to its current value.
   The much simpler closed box model of chemical evolution presented in \citet{DH1998}
   predicts a gradual increase in star formation and abundances
   over time. The third model is an AMR derived by \citet[][hereafter the HZ04, model]{HZ2004}
   based on UBVI photometry from the Magellanic Clouds Photometric Survey. The authors'
   conclusions are consistent with the PT98 model. They surmised that approximately
   50\% of the stars in the SMC formed prior to 8.4~Gyr, and that between 8.4 and 3~Gyr
   ago   star formation was not efficient. The next rise in the mean star formation
   rate occurred during recent 3~Gyr with bursts at ages of 2.5~Gyr, 400~Myr,
   and 60~Myr, where the first two bumps are consistent with past  perigalactic passages
   by the SMC with the Milky Way. The major merger scenario for the SMC was proposed
   by \citet[][hereafter the TB09 model]{TB2009}. It predicts that a~major merger occurred
   $\sim$7.5~Gyr ago and was calculated for three cases: no merger,  TB09-1;
   one-to-one merger,  TB09-2; and one-to-four merger,  TB09-3. A~quite different
   formation history is proposed by \citet{Cignoni2013} through the  two  Bologna models,
   (C13-B) and Cole (C13-C). They predict fast initial enrichment prior to 9~Gyr
   ago, then monotonic increase in metallicity between 9 to 4~Gyr with no evidence
   of metallicity dips, followed by another enrichment at more recent times.

   Our AMR presented in Fig.~\ref{fig:am} in general is very consistent with the
   models. The older clusters Lindsay~19, 27, and 113 closely follow  the burst of star
   formation at $\sim$3~Gyr ago from the PT98 model. NGC~339 and Lindsay~1 are
   consistent with the PT98 model within the errors, but more closely reflect the  TB09-1
   or C13-C model suggesting the existence of burst at about 7~Gyr. NGC~361 and Lindsay~6
   seem to represent the closed box model the best. Both these clusters
   have relatively high reddening according to reddening maps, but we adopted much
   smaller values for the metallicity calculation.
   This can mean that their position with respect to the SMC might be
   different from the distance we use   in the present study. However, a~different
   distance would change the age, although not the metallicity. Another possibility
   is that these clusters are located in regions richer in gas and could be chemically
   enriched during their formation or evolution.
   The HZ04 model is on average too metal rich for the discussed clusters, which
   is also true for other literature values derived from photometric or spectroscopic
   data (see Fig.~\ref{fig:amlit}). Only values provided by \citet{Perren2017}
   closely follow  this model at ages older than 2.5~Gyr.
   \citet{DH1998} described Lindsay~113 and NGC~339 as anomalous, and
   suggested that they could have been formed from the infall of unenriched,
   or  less enriched, gas. However, \citet{Parisi2009} showed that these two
   clusters behave well, but in turn Lindsay~1 appears too metal rich with respect
   to the PT98 model. We do not see any anomalous behavior of Lindsay~113 or NGC~339,
   but Lindsay~1 indeed seems to be too metal rich. There is a~discrepancy between
   the ages we determined based on the Dartmouth and Padova databases. For clusters older
   than about 3~Gyr the Dartmouth isochrones give older ages than the Padova isochrones,
   and for clusters younger than 3~Gyr the effect is the opposite.
   Moreover, it seems that the older the cluster is, the larger the age difference
   becomes for these two sets of isochrones.

   Clusters younger than 1~Gyr in our sample fit equally well all models
   of chemical evolution, although they appear to be systematically less metal rich
   than predictions. The results for these clusters are less reliable due to small
   number of stars used for the metallicity calculation. \citet{Parisi2009} indicated
   that their metallicity for NGC~330 is significantly more metal poor than the PT98
   model prediction. This is also true for our case. NGC~330 deviates greatly from the AMR,
   although the number of stars used for its metallicity measurement is sufficient
    for a~reliable mean metallicity determination. Furthermore, the metallicity
   for NGC~376 is extremely low in our AMR. This result however is based
   on only one star in the first field, and two stars in the second field.
   Apart from these two outliers, other young star clusters are still characterized
   by quite large spread in the metallicity. However, overall compliance with the
   theoretical predictions is satisfactory.

\section{Discussion}
\label{sec:discussion}

  Figure~\ref{fig:amlit} shows our AMR with superimposed literature values from
  Table~\ref{tab:lit}: metallicities derived from spectroscopy (low- and high-resolution),
  Str\"omgren photometry. and RGB slope method all expressed on the ZW84 scale, as
  well as values obtained by Perren et al. (2017) with the ASteCA package, and from
  theoretical isochrone fitting obtained by various authors. Figure~\ref{fig:fehlit}
  presents the comparison between the derived metallicities and their literature
  counterparts.

   The metallicities and ages of intermediate-age star clusters in our sample are
   on average more metal rich and, consequently, younger than most literature values
   expressed on the same metallicity scale. The possible explanation  is the adopted
   calibration of Str\"omgren colors with metallicity.
   Data points based on Str\"omgren photometry from \citet{Livanou2013},
   \citet{Piatti2018}, and \citet{Piatti2019} were obtained with different calibrations
   \citep[][respectively]{HRG1995, GR1992, Calamida2007}.
   \citet{Hilker2000} presented an extended calibration of \citet{GR1992} for cluster
   and field red giants for a~wide range of metallicities ($-2.0 < \mathrm{[Fe/H]} < 0.0$~dex).
   The original calibration yields higher metallicities for redder stars and lower
   metallicities for bluer stars in the color range used for metallicity determination.
   Older clusters and fields have many stars in the color range of $0.5-0.7$~mag,
   consequently their mean metallicities calculated based on the \citet{GR1992} calibration
   would be shifted toward more metal-poor values than the \citet{Hilker2000} calibration.
   On the other hand, metallicities calculated based on the \citet{Calamida2007}
   calibration, derived based on red giants from Galactic globular clusters in the
   metallicity range of $-2.2 < \mathrm{[Fe/H]} < -0.7$~dex, gives systematically
   more metal-poor results than \citet[][see Table~6 in their work]{Hilker2000}.
   Nevertheless, we decided to use the \citet{Hilker2000} calibration in our study
   as it is calibrated for the widest range of metallicities, and it can be used
   for both old metal-poor and young metal-rich giants and supergiants.
   This choice motivated us to reanalyze the data sets published by \citet{Piatti2018}
   and \citet{Piatti2019}. To keep the uniformity in the analysis, we redid the
   photometry and calibrated the data using direct color--color transformation equations
   for the two chips separately, which was an improvement as the zero points of the
   two chips of the camera differ slightly.
   For the selection of the cluster members we used the updated catalog of \citet{Bica2020}
   and the applied additional step of rejecting foreground stars based on their
   Gaia proper motions. For most of the overlapping clusters (except two) we adopted
   different reddening values from new reddening maps of G20 and S21. These changes
   resulted in the selection of different stars for metallicity calculation, which is
   especially visible in the case of young clusters. The resulting discrepancies between
   our findings and other studies based on Str\"omgren photometry are the smallest
   for metal-poor clusters ($\sim$0.1~dex) and much larger ($\sim$0.4~dex) for
   clusters more metal rich than $-0.70$~dex.

   The metallicities from low-resolution spectroscopy from \citet{DH1998} are
   significantly more metal poor than our results (by $\sim$0.26~dex), and consequently
   the clusters are older.
   Interestingly, the metallicities of NGC~330 from high-resolution spectroscopy
   agree very well with our determination \citep[($\Delta \mathrm{[Fe/H]} \approx
   -0.06$~dex);][]{Spite1991,Hill1997,GW1999}.
   Data points from the RGB slope method from \citet{Mighell1998} are also systematically
   more metal poor than our results (by $\sim$0.36~dex). Spectroscopic and RGB slope
   points seem to follow a~metallicity dip predicted by the TB09-3 model. Our results
   do not show such a~behavior.
   The data points from \citet{Perren2017} obtained with the ASteCA package are systematically
   more metal rich than any other method (in our case $\Delta \mathrm{[Fe/H]} \approx
   -0.33$~dex). The overall agreement of our results with literature values from
   isochrones fitting and Washington photometry are satisfactory ($\Delta \mathrm{[Fe/H]}
   \approx -0.11$~dex and 0.09~dex, respectively).

   The conclusion of \citet{Livanou2013} was that there is no indication of an AMR
   in the SMC. We do not confirm that statement. Despite of a handful of low-metallicity
   young star clusters like NGC~330, NGC~376, and OGLE-CL~SMC~68 and  69, the average
   metallicity of young clusters is $-0.70$~dex with $\sigma = 0.22$~dex, while for
   older ones it is $-0.99$~dex with $\sigma = 0.13$~dex.

   \citet{Piatti2007a} noticed the tendency that the clusters located in the inner
   regions of the SMC are younger than  those from outer regions, while their
   mean metallicity and its dispersion are greater close to the SMC center. We confirm
   both of these observations (see Fig.~\ref{fig:mapfeh}).
   Furthermore, \citet{Piatti2012} pointed out that   stellar populations younger
   than about 2~Gyr are more metal rich than $\mathrm{[Fe/H] \approx -0.8}$~dex
   and are located in the innermost region confined to an ellipse with a~semi-major
   axis $\lesssim 1^{\circ}$. We find that on average the younger stellar populations
   from this region have metallicities of about or less than $-0.8$~dex, although
   there are single fields that  break this rule (e.g., NGC~330).
   \citet{Piatti2012} found that with increasing semi-major axis the field stars
   became older and more metal poor. We confirm this trend. The AMR presented by
   \citet{Piatti2012} for the field stars shows two bumps indicating enhanced formation
   processes at about 2 and 7.5~Gyr. The three oldest clusters in our sample seem
   to reflect such a~bump at about 7~Gyr.

   \citet{Parisi2009}, and later \citet{Parisi2014}, indicated that the PT98
   model represents  the AMR well  for ages $< 3-4$~Gyr, but fails for older ages.
   In their AMR star clusters younger than $\sim$4~Gyr follow closely the burst from
   the PT98 model, except three clusters fitting better the closed box model
   of \citet{DH1998}.
   Very similar behavior is also visible in our AMR. Furthermore, the three oldest
   clusters from their sample seem to reproduce a~burst at about 7.5~Gyr not present
   in the PT98 model, but predicted by the TB09 set of models. In our AMR the oldest
   clusters are also represented better by TB09 models.

   In the conclusions \citet{Perren2017} indicated that the metallicities obtained
   with the ASteCA package are on average $\sim$0.22~dex higher than literature values.
   In addition, their AMR cannot be successfully matched by any model or exiting empirical
   determination. We confirm this statement as our findings are also systematically
   more metal poor than the values from \citet{Perren2017}.

\section{Summary and conclusions}
\label{sec:summ}

   In this work we presented the analysis of Str\"omgren photometry of 35 star
   clusters from the SMC in order to obtain their mean metallicities and ages.
   We also provided mean metallicities of the fields surrounding the clusters.
   Metallicities and ages were derived in a~consistent manner by using the relation
   of photometric metallicity and Str\"omgren colors calibrated by \citet{Hilker2000},
   which allowed us to compare the obtained results and trace the metallicity and
   age distribution across the SMC. Moreover, we used for the calculations the
   most recent reddening maps of G20 and S21, as well as the distance to the SMC
   derived by \citet{Graczyk2020}, which is precise to 2\%.

   The metallicity distribution of the field stars in the SMC shows a~trend typical
   of irregular galaxies. The more metal-rich stars tend to accumulate close to the
   central region of the SMC. The farther away from the SMC main body, the more
   metal poor the stars become. The average metallicity values of the young field giants and
   supergiants ($-0.70$~dex) and old field stars ($-0.73$~dex) are similar within
   the errors. The average metallicity of HBG stars found by us is in close agreement
   with the results reported by \citet[][$-0.60$~dex]{Ripepi2017},
   \citet[][$-0.75$~dex]{Romaniello2008}, or \citet[][$-0.74$~dex]{Lemasle2017}
   for Classical Cepheids, a subset of such stars.
   The age distribution of clusters in the SMC confirms earlier studies by
   \citet{Carrera2008} or \citet{Glatt2010}, among others,  that young stellar clusters distribute
   along the SMC main body while the intermediate-age clusters are located farther
   from it.

   The two features described above are reflected in the AMR constructed for studied
   stellar clusters.
   The majority of clusters analyzed in this paper are young, with very few stars
   for reliable metallicity and age calculation. Only seven clusters in our sample
   have well populated RGBs, where the majority of stars for which Str\"omgren
   photometry provides metallicity estimates are found. The overall results agree
   well with theoretical models of chemical enrichment of the SMC, and with
   previous literature studies. The metallicities for seven star clusters
   (OGLE-CL~SMC~68, 71, 82, 88, 126,  143, and Bruck~39) are provided for the first
   time, according to our knowledge.
   Three intermediate-age clusters (Lindsay~19, 27, and 113) reproduce well the
   burst of chemical enrichment $\sim$3~Gyr in the PT98 model, while NGC~361 appears
   to be too metal rich. The three oldest clusters (NGC~339, Lindsay~1 and 6) seem
   to follow the burst from the TB09 models (predicted at $\sim$7.5~Gyr).
   The spread of metallicities for young clusters is quite large, but shows no indication
   of any trend.
   We provide the catalog of Str\"omgren photometry in the SMC available through the webpage
   of the Araucaria Project and the CDS.

   This study proves the usefulness of the Str\"omgren filters in stellar astrophysics and
   shows the potential of this photometric system in the future studies of   single stars
   and of large groups of stars like stellar clusters or galaxies.

\begin{acknowledgements}
   We thank the anonymous referee for valuable comments which improved this paper.

   The research leading to these results has received funding from the European Research
   Council (ERC) under the European Union’s Horizon 2020 research and innovation program
   (grant agreement No 695099). We also acknowledge support from the National Science Center,
   Poland grants MAESTRO UMO-2017/26/A/ST9/00446, BEETHOVEN UMO-2018/31/G/ST9/03050 and
   DIR/WK/2018/09 grants of the Polish Ministry of Science and Higher Education.

   We gratefully acknowledge financial support for this work from the BASAL Centro de
   Astrofisica y Tecnologias Afines (CATA) AFB-170002 and the Millenium Institute
   of Astrophysics (MAS) of the Iniciativa Cientifica Milenio del Ministerio de Economia,
   Fomento y Turismo de Chile, project IC120009.

   M.G. gratefully acknowledges support from FONDECYT POST- DOCTORADO grant 3170703.

\end{acknowledgements}

\begin{sidewaystable*}
\caption{Star clusters in the SMC. Cluster: name of the cluster;
RA, DEC: equatorial coordinates of the cluster for epoch J2000 from Bica et al. (2020);
Date: date of observation;
T$_{\mathrm{exp}}$:  exposure time of filter $y$, $b$, and $v$; Airmass:  airmass of observations;
Seeing: average seeing; Other name: other name of the clusters in use.}
\label{tab:smc}
\centering
\begin{tabular}{llllrlll}
\hline\hline
Cluster & RA & DEC & Date & T$_{exp}$ ($y$,$b$,$v$) & Airmass ($y$,$b$,$v$)
& Seeing ($y$,$b$,$v$) & Other name  \\
 & (hh:mm:ss.s) & (dd:mm:ss.s) & & (s) &  & (arcsec) &\\
\hline
Lindsay~113 & 01:49:30.3 & -73:43:40 & 2008~Dec~17 & 120;200;500 & 1.46;1.47;1.48
& 0.72;0.77;0.62 & ESO~30-4 \\
NGC~339     & 00:57:47.5 & -74:28:17 & 2008~Dec~17 & 180;300;500 & 1.61;1.62;1.63
& 0.88;0.86;0.69 & Lindsay~59, ESO~29-25, Kron~36\\
NGC~361     & 01:02:11.0 & -71:36:21 & 2008~Dec~17 & 180;300;500 & 1.61;1.62;1.64
& 0.82;0.88;0.86 & Lindsay~67, ESO~51-12, Kron~46 \\
Lindsay~1   & 00:03:54.6 & -73:28:16 & 2008~Dec~17 & 120;200;500 & 1.88;1.89;1.90
& 0.84;0.88;0.62 & ESO~28-8 \\
Lindsay~6   & 00:23:04.1 & -73:40:12 & 2008~Dec~17 & 180;300;500 & 1.88;1.89;1.91
& 0.82;0.77;0.83 & ESO~28-17, Kron~4 \\
NGC~330     & 00:56:18.7 & -72:27:48 & 2008~Dec~17 & 120;200;400 & 1.49;1.49;1.50
& 0.72;0.77;0.83 & ESO~29-24, Kron~35, Lindsay~54, \\
            &  &  & 2008~Dec~19 & 100,160;400 & 1.47;1.48;1.48 & 0.89;0.96;0.99 & OGLE-CL~SMC~107 \\
OGLE-CL~SMC~45  & 00:48:01.0 & -73:29:10 & 2008~Dec~18 & 120;200;500 & 1.81;1.81;1.83
& 1.29;1.31;1.17 & Lindsay~35, Kron~25 \\
Lindsay~27  & 00:41:24.2 & -72:53:27 & 2008~Dec~18 & 180;300;500 & 1.44;1.45;1.46
& 1.05;1.12;1.34 & Kron~21, OGLE-CL~SMC~12 \\
Lindsay~19  & 00:37:41.8 & -73:54:27 & 2008~Dec~18 & 180;300;500 & 1.50;1.50;1.51
& 1.22;1.11;1.23 & OGLE-CL~SMC~3 \\
NGC~265     & 00:47:11.4 & -73:28:37 & 2008~Dec~18 & 180;300;500 & 1.51;1.51;1.52
& 1.12;1.06;1.16 & Lindsay~34, OGLE-CL~SMC~39 \\
            & & & 2008~Dec~18 & 120;200;500 & 1.74;1.74;1.77
            & 1.16;1.26;1.29 & ESO~29-14, Kron~24  \\
NGC~376     & 01:03:53.7 & -72:49:32 & 2008~Dec~18 & 180;300;500 & 1.52;1.52;1.53
& 0.94;1.23;1.31 & Lindsay~72, ESO~29-29, Kron~49 \\
            & & & 2008~Dec~19 & 90;140;350 & 1.52;1.52;1.53
            & 0.83;0.92;0.95 & OGLE-CL~SMC~139 \\
IC~1611     & 00:59:47.7 & -72:20:02 & 2008~Dec~18 & 120;200;500 & 1.56;1.56;1.57
& 1.16;1.09;1.12 & Lindsay~61, ESO~29-27, Kron~40 \\
            &  &  &  &  &  &  & OGLE-CL~SMC~118 \\
IC~1612     & 01:00:02.1 & -72:22:05 & 2008~Dec~18 & 120;200;500 & 1.56;1.56;1.57
            & 1.16;1.09;1.12 & Lindsay~62, ESO~29-28, Kron~41 \\
            &  &  &  &  &  &  & OGLE-CL~SMC~120 \\
OGLE-CL~SMC~32  & 00:45:54.1 & -73:30:24 & 2008 Dec 18 & 120;200;500 & 1.69;1.70;1.71
& 1.32;1.29;1.32 & NGC~256, Lindsay~30, ESO~29-11 \\
                &  &  &  &  &  &  & Kron~23 \\
Bruck~39    & 00:45:27.1 & -73:28:50 & 2008 Dec 18 & 120;200;500 & 1.69;1.70;1.71
& 1.32;1.29;1.32 & OGLE-CL~SMC~27 \\
OGLE-CL~SMC~68  & 00:50:56.9 & -73:17:15 & 2008~Dec~19 & 120;200;500 & 1.43;1.43;1.43
& 0.84;0.94;0.94 & [BS95]~40 \\
OGLE-CL~SMC~69  & 00:51:14.9 & -73:09:40 & 2008~Dec~19 & 100;160;400 & 1.45;1.45;1.45
& 0.83;0.83;0.92 & NGC~290, Lindsay~42, ESO~29-19 \\
OGLE-CL~SMC~99  & 00:54:47.5 & -72:27:57 & 2008~Dec~19 & 100;160;400 & 1.46;1.46;1.46
& 0.79;0.96;1.06 & Bruck~79 \\
OGLE-CL~SMC~129 & 01:01:44.6 & -72:33:51 & 2008~Dec~19 & 100;160;400 & 1.49;1.50;1.50
& 0.77;0.77;0.85 & Lindsay~66 \\
OGLE-CL~SMC~142 & 01:04:36.1 & -72:09:39 & 2008~Dec~19 & 90;140;350 & 1.53;1.54;1.54
& 0.90;0.91;1.01 & Lindsay~74, ESO~51-15, Kron~50 \\
$\mathrm{[BS95]}$~123 & 01:04:27.6 & -72:10:55 & 2008~Dec~19 & 90;140;350 & 1.53;1.54;1.54
& 0.90;0.91;1.01 & \\
OGLE-CL~SMC~144 & 01:04:05.2 & -72:07:14 & 2008~Dec~19 & 90;140;350 & 1.53;1.54;1.54
& 0.90;0.91;1.01 & OGLE-CL~SMC~236 \\
OGLE-CL~SMC~49  & 00:48:37.0 & -73:24:53 & 2008~Dec~19 & 90;140;350 & 1.65;1.66;1.66
& 0.88;0.97;0.95 & Bruck~48 \\
OGLE-CL~SMC~50  & 00:49:00.0 & -73:09:05 & 2008~Dec~19 & 90;140;350 & 1.68;1.69;1.70
& 0.80;0.84;0.99 & \\
OGLE-CL~SMC~54  & 00:49:03.1 & -73:21:40 & 2008~Dec~19 & 90;140;350 & 1.73;1.74;1.75
& 0.88;0.93;0.91 & Lindsay~39 \\
OGLE-CL~SMC~61  & 00:50:02.0 & -73:15:24 & 2008~Dec~19 & 90;140;350 & 1.82;1.83;1.84
& 0.85;0.93;0.94 & [H86]~107 \\
OGLE-CL~SMC~71  & 00:51:32.2 & -73:00:48 & 2008~Dec~19 & 90;140;350 & 1.87;1.87;1.88
& 0.82;0.87;0.85 & Bruck~57 \\
OGLE-CL~SMC~205 & 00:51:31.8 & -72:58:44 & 2008~Dec~19 & 90;140;350 & 1.87;1.87;1.88
& 0.82;0.87;0.85 & [H86]~124 \\
OGLE-CL~SMC~78  & 00:52:16.7 & -73:01:04 & 2009~Jan~16 & 120;200;500 & 1.65;1.64;1.66
& 0.98;1.02;0.95 & [H86]~130 \\
OGLE-CL~SMC~82  & 00:52:42.6 & -72:55:30 & 2009~Jan~16 & 90;140;380 & 1.73;1.72;1.70
& 0.83;0.85;0.91 & [BS95]~60 \\
OGLE-CL~SMC~88  & 00:53:01.0 & -72:53:49 & 2009~Jan~16 & 90;140;380 & 1.73;1.72;1.70
& 0.83;0.85;0.91 & Lindsay~46, Kron~31 \\
OGLE-CL~SMC~126 & 01:01:00.7 & -72:45:00 & 2009~Jan~16 & 90;140;350 & 1.72;1.73;1.74
& 0.87;0.92;1.06 & Lindsay~65, [H86]~192 \\
OGLE-CL~SMC~128 & 01:01:37.0 & -72:24:25 & 2009~Jan~17 & 100;180;380 & 1.56;1.57;1.57
& 0.90;0.91;1.07 & Bruck~105 \\
OGLE-CL SMC 143 & 01:04:40.6 & -72:33:03 & 2009~Jan~17 & 90;140;350 & 1.61;1.61;1.59
& 1.27;1.25;1.14 & [BS95]~125 \\
OGLE-CL~SMC~156 & 01:07:28.0 & -72:46:10 & 2009~Jan~18 & 120;200;750 & 1.62;1.61;1.58
& 1.24;1.17;1.19 & Lindsay~80 \\
\hline
\end{tabular}
\end{sidewaystable*}

%
\begin{table*}
\caption{Transformation coefficients.}
\label{tab:std}
\centering
\begin{tabular}{ccrrrrcc}     
\hline\hline
Night & chip & eq. & coeff$_1$ & coeff$_2$ & coeff$_3$ & coeff$_4$ & r.m.s.\\
\hline
 & & $y$ & 1.164 $\pm$ 0.013 & 0.000 $\pm$ 0.028 & 0.102 $\pm$ 0.018 & - & 0.015 \\
 & 1 & $(b-y)$ & 0.059 $\pm$ 0.010 & 0.985 $\pm$ 0.021 & 0.060 $\pm$ 0.014 & - & 0.011 \\
 17 Dec & & $m1$ & 0.352 $\pm$ 0.013 & 0.125 $\pm$ 0.059 & 0.043 $\pm$ 0.017 & 0.915 $\pm$ 0.073 & 0.011 \\
 2008 & & $y$ & 1.203 $\pm$ 0.006 & -0.000 $\pm$ 0.014 & 0.096 $\pm$ 0.009 & - & 0.007 \\
 & 2 & $(b-y)$ & 0.059 $\pm$ 0.008 & 0.963 $\pm$ 0.018 & 0.041 $\pm$ 0.011 & - & 0.010 \\
 & & $m1$ & 0.295 $\pm$ 0.008 & 0.206 $\pm$ 0.038 & 0.098 $\pm$ 0.013 & 0.873 $\pm$ 0.045 & 0.009 \\ \hline

 & & $y$ & 1.126 $\pm$ 0.005 & -0.026 $\pm$ 0.012 & 0.094 $\pm$ 0.010 & - & 0.007 \\
 & 1 & $(b-y)$ & 0.062 $\pm$ 0.009 & 0.974 $\pm$ 0.019 & 0.058 $\pm$ 0.017 & - & 0.010 \\
 18 Dec & & $m1$ & 0.331 $\pm$ 0.008 & 0.117 $\pm$ 0.033 & 0.055 $\pm$ 0.016 & 0.978 $\pm$ 0.037 & 0.008 \\
 2008 & & $y$ & 1.162 $\pm$ 0.005 & -0.046 $\pm$ 0.012 & 0.011 $\pm$ 0.009 & - & 0.005 \\
 & 2 & $(b-y)$ & 0.039 $\pm$ 0.007 & 1.003 $\pm$ 0.017 & 0.066 $\pm$ 0.013 & - & 0.008 \\
 & & $m1$ & 0.323 $\pm$ 0.010 & 0.108 $\pm$ 0.049 & 0.044 $\pm$ 0.019 & 0.971 $\pm$ 0.051 & 0.009 \\ \hline

 & & $y$ & 1.093 $\pm$ 0.019 & 0.004 $\pm$ 0.034 & 0.108 $\pm$ 0.026 & - & 0.023 \\
 & 1 & $(b-y)$ & 0.063 $\pm$ 0.004 & 0.966 $\pm$ 0.008 & 0.049 $\pm$ 0.005 & - & 0.005 \\
 19 Dec & & $m1$ & 0.323 $\pm$ 0.012 & 0.054 $\pm$ 0.058 & 0.063 $\pm$ 0.017 & 1.082 $\pm$ 0.070 & 0.012 \\
 2008 & & $y$ & 1.118 $\pm$ 0.008 & -0.001 $\pm$ 0.016 & 0.130 $\pm$ 0.011 & - & 0.011 \\
 & 2 & $(b-y)$ & 0.042 $\pm$ 0.005 & 0.983 $\pm$ 0.009 & 0.045 $\pm$ 0.006 & - & 0.006 \\
 & & $m1$ & 0.319 $\pm$ 0.012 & 0.070 $\pm$ 0.051 & 0.061 $\pm$ 0.016 & 1.041 $\pm$ 0.060 & 0.014 \\ \hline

 & & $y$ & 1.152 $\pm$ 0.008 & -0.001 $\pm$ 0.017 & 0.095 $\pm$ 0.012 & - & 0.008 \\
 & 1 & $(b-y)$ & 0.073 $\pm$ 0.007 & 0.966 $\pm$ 0.016 & 0.066 $\pm$ 0.013 & - & 0.008 \\
 16 Jan & & $m1$ & 0.347 $\pm$ 0.009 & -0.012 $\pm$ 0.050 & 0.047 $\pm$ 0.015 & 1.117 $\pm$ 0.055 & 0.009\\
 2009 & & $y$ & 1.206 $\pm$ 0.006 & -0.031 $\pm$ 0.015 & 0.104 $\pm$ 0.012 & - & 0.005 \\
 & 2 & $(b-y)$ & 0.072 $\pm$ 0.009 & 0.972 $\pm$ 0.021 & 0.045 $\pm$ 0.018 & - & 0.008 \\
 & & $m1$ & 0.288 $\pm$ 0.011 & -0.001 $\pm$ 0.064 & 0.121 $\pm$ 0.026 & 1.149 $\pm$ 0.086 & 0.008\\ \hline

 & & $y$ & 1.176 $\pm$ 0.018 & 0.036 $\pm$ 0.039 & 0.149 $\pm$ 0.013 & - & 0.015 \\
 & 1 & $(b-y)$ & 0.060 $\pm$ 0.008 & 0.996 $\pm$ 0.018 & 0.046 $\pm$ 0.006 & - & 0.008 \\
 17 Jan & & $m1$ & 0.345 $\pm$ 0.020 & 0.113 $\pm$ 0.151 & 0.077 $\pm$ 0.020 & 0.925 $\pm$ 0.200 & 0.016\\
 2009 & & $y$ & 1.231 $\pm$ 0.008 & 0.033 $\pm$ 0.016 & 0.147 $\pm$ 0.005 & - & 0.006 \\
 & 2 & $(b-y)$ & 0.069 $\pm$ 0.011 & 0.962 $\pm$ 0.025 & 0.058 $\pm$ 0.009 & - & 0.011 \\
 & & $m1$ & 0.290 $\pm$ 0.014 & -0.049 $\pm$ 0.111 & 0.038 $\pm$ 0.015 & 1.209 $\pm$ 0.141 & 0.013\\ \hline

 & & $y$ & 1.222 $\pm$ 0.015 & -0.051 $\pm$ 0.033 & 0.131 $\pm$ 0.012 & - & 0.016 \\
 & 1 & $(b-y)$ & 0.072 $\pm$ 0.013 & 0.988 $\pm$ 0.029 & 0.045 $\pm$ 0.011 & - & 0.013 \\
 18 Jan & & $m1$ & 0.321 $\pm$ 0.016 & 0.118 $\pm$ 0.087 & 0.054 $\pm$ 0.022 & 0.999 $\pm$ 0.095 & 0.012\\
 2009 & & $y$ & 1.257 $\pm$ 0.009 & -0.024 $\pm$ 0.019 & 0.125 $\pm$ 0.006 & - & 0.008 \\
 & 2 & $(b-y)$ & 0.077 $\pm$ 0.010 & 0.988 $\pm$ 0.021 & 0.036 $\pm$ 0.007 & - & 0.009 \\
 & & $m1$ & 0.288 $\pm$ 0.017 & 0.132 $\pm$ 0.088 & 0.078 $\pm$ 0.017 & 0.935 $\pm$ 0.088 & 0.010\\
\hline
\end{tabular}
\end{table*}
%

%
\begin{table*}
\caption{Star clusters in the SMC.
Cluster: name of the cluster;
D$_{maj}$;D$_{min}$:  major and minor axes from the updated catalog of \citet{Bica2020};
$E(B-V)_C$:  reddening adopted for the cluster stars;
$\mathrm{[Fe/H]}_C$: mean cluster metallicity calculated in this work (systematic errors
are given in the brackets);
N$_C$: number of stars used for metallicity calculation;
log(Age$_P$): logarithm of age derived from the Padova isochrones;
log(Age$_D$): logarithm of age derived from the Dartmouth isochrones.}
\label{tab:reddC}
\centering
\begin{tabular}{lclcccc}     
\hline\hline
Cluster & D$_{maj}$;D$_{min}$ & $E(B-V)_C$ & $\mathrm{[Fe/H]}_C$ & N$_C$
& log(Age$_{P}$) & log(Age$_{D}$)\\
 & (arcmin) & (mag) & (dex) & & (yr) & (yr) \\
\hline
Lindsay~113   & 4.4;4.4 & 0.03  & -1.14$\pm$0.03 (0.10) & 57 & 9.55$\pm$0.04 & 9.60$\pm$0.03 \\
NGC~339       & 2.9;2.9 & 0.041 & -1.10$\pm$0.03 (0.12) & 93 & 9.75$\pm$0.08
& 9.78$^{+0.07}_{-0.08}$ \\
NGC~361       & 2.6;2.6 & 0.03  & -0.79$\pm$0.04 (0.12) & 81 & 9.51$\pm$0.10
& 9.54$^{+0.11}_{-0.15}$ \\
Lindsay~1     & 4.6;4.6 & 0.033 & -1.06$\pm$0.03 (0.10) & 61 & 9.82$\pm$0.06
& 9.85$^{+0.06}_{-0.07}$ \\
Lindsay~6     & 1.7;1.7 & 0.03  & -0.91$\pm$0.07 (0.12) & 18 & 9.80$\pm$0.07
& 9.85$^{+0.06}_{-0.07}$ \\
NGC~330       & 2.8;2.5 & 0.075 & -0.98$\pm$0.08 (0.10) & 9  & 7.50$\pm$0.10 & - \\
              &  &  & -0.98$\pm$0.07 (0.09)$^{(1)}$ & 8 &  &  \\
OGLE-CL~SMC~45  & 1.2;1.2 & 0.063 & -0.31$\pm$0.10 (0.13) & 6 & 8.31$\pm$0.15 & - \\
Lindsay~27    & 2.5;2.5 & 0.068 & -1.04$\pm$0.03 (0.10) & 43 & 9.50$\pm$0.08
& 9.54$^{+0.11}_{-0.15}$ \\
Lindsay~19    & 1.7;1.7 & 0.062 & -0.86$\pm$0.05 (0.11) & 25 & 9.42$\pm$0.12
& 9.40$^{+0.15}_{-0.22}$ \\
NGC~265       & 1.2;1.2 & 0.075 & -0.75$\pm$0.08 (0.11) & 11 & 8.53$\pm$0.10 & - \\
            & & & -0.79$\pm$0.09 (0.12)$^{(1)}$ & 12 &  &  \\
NGC~376     & 1.8;1.8 & 0.067 & -1.20$\pm$0.28 (0.09)$^{(1)}$ & 1 &  &  \\
            & & & -0.98$\pm$0.21 (0.09) & 2 & 7.67$\pm$0.20 & - \\
IC~1611     & 1.5;1.5 & 0.03  & -0.58$\pm$0.09 (0.12) & 7 & 8.20$\pm$0.10 & - \\
IC~1612     & 1.2;0.8 & 0.085 & -0.68$\pm$0.08 (0.10) & 3 & 7.69$\pm$0.08 & - \\
OGLE-CL~SMC~32  & 0.9;0.9 & 0.083 & -0.33$\pm$0.31 (0.12) & 1 & 7.96$\pm$0.07 & - \\
Bruck~39    & 0.55;0.55 & 0.083 & -0.57$\pm$0.14 (0.11) & 3 & 8.58$\pm$0.13 & - \\
OGLE-CL~SMC~68$^{(*)}$  & 1.1;0.8 & 0.060 & -0.97$\pm$0.10 (0.10) & 2 & 7.74$\pm$0.14 & - \\
OGLE-CL~SMC~69  & 1.1;1.1 & 0.074 & -0.86$\pm$0.13 (0.10) & 4 & 7.97$\pm$0.10 & - \\
OGLE-CL~SMC~99  & 0.8;0.65 & 0.052 & - & - & 7.65$\pm$0.03 & - \\
OGLE-CL~SMC~129 & 1.1;1.1 & 0.089 & - & - & 7.23$\pm$0.05 & - \\
OGLE-CL~SMC~142 & 1.0;1.0 & 0.070 & - & - & 7.05$\pm$0.10 & - \\
$\mathrm{[BS95]}$~123$^{(*)}$  & 1.1;0.85 & 0.070 & -0.69$\pm$0.09 (0.09) & 2 & 7.45$\pm$0.15 & - \\
OGLE-CL~SMC~144 & 0.6;0.6 & 0.070 & - & - & 7.64$\pm$0.05 & - \\
OGLE-CL~SMC~49  & 1.3;1.1 & 0.065 & - & - & 8.17$\pm$0.05 & - \\
OGLE-CL~SMC~50  & 0.75;0.75 & 0.123 & - & - & 8.27$\pm$0.05 & - \\
OGLE-CL~SMC~54  & 0.70.55 & 0.058 & -0.84$\pm$0.05 (0.11) & 3 & 8.09$\pm$0.20 & - \\
OGLE-CL~SMC~61  & 1.1;0.8 & 0.082 & - & - & 7.74$\pm$0.05 & - \\
OGLE-CL~SMC~71$^{(*)}$  & 0.45;0.45 & 0.062 & -1.01$\pm$0.14 (0.10) & 2 & 8.25$\pm$0.10 & - \\
OGLE-CL~SMC~205  & 0.85;0.65 & 0.062 & - & - & 8.20$\pm$0.12 & - \\
OGLE-CL~SMC~78   & 0.75;0.60 & 0.046 & - & - & 7.84$\pm$0.20 & - \\
OGLE-CL~SMC~82$^{(*)}$  & 0.8;0.8 & 0.057 & -0.70$\pm$0.33 (0.11) & 1 & 8.17$\pm$0.10 & - \\
OGLE-CL~SMC~88$^{(*)}$  & 4.3;4.3 & 0.057 & -0.65$\pm$0.05 (0.10) & 5 & 7.85$\pm$0.07 & - \\
OGLE-CL~SMC~126$^{(*)}$ & 1.1;1.1 & 0.063 & -0.78$\pm$0.31 (0.10) & 1 & 7.99$\pm$0.08 & - \\
OGLE-CL~SMC~128 & 0.75;0.75 & 0.087 & - & - & 7.30$\pm$0.30 & - \\
OGLE-CL~SMC~143$^{(*)}$ & 1.60;0.85 & 0.089 & -0.45$\pm$0.04 (0.12) & 3 & 8.03$\pm$0.20 & - \\
OGLE-CL~SMC~156 & 1.2;1.2 & 0.068 & -0.55$\pm$0.18 (0.12) & 2 & 8.16$\pm$0.10 & - \\
\hline
\end{tabular}
\tablefoot{
\begin{itemize}\scriptsize
\itemsep0em
\item [$^{(*)}$] Metallicities of the clusters provided for the first time.
\item [$^{(1)}$] Not used for the AMR because of the lower quality.
\end{itemize}}
\end{table*}
%

%
\begin{table*}
\caption{Fields surrounding star clusters in the SMC.
Field: name of the cluster in the field;
$E(B-V)_F$: Reddening adopted for the field stars;
$\mathrm{[Fe/H]}_F$: mean metallicity of the field stars (systematic errors are given
in the brackets);
$\mathrm{[Fe/H]}_{bF}$: mean metallicity of the young field giants;
N$_F$, N$_{bF}$: number of stars used for the mean metallicity calculation of old and young
giants, respectively.}
\label{tab:reddF}
\centering
\begin{tabular}{lccccc}     
\hline\hline
Field & $E(B-V)_F$ & $\mathrm{[Fe/H]}_F$ & N$_F$ & $\mathrm{[Fe/H]}_{bF}$
& N$_{bF}$\\
 & (mag) & (dex) & & (dex) & \\
\hline
Lindsay~113   & 0.058 & - & - & - & - \\
NGC~339       & 0.041 & -1.05$\pm$0.03 (0.10) & 95 & - & - \\
NGC~361       & 0.080 & -0.85$\pm$0.04 (0.10) & 58 & - & - \\
Lindsay~1     & 0.033 & -1.07$\pm$0.04 (0.11) & 5 & - & -  \\
Lindsay~6     & 0.048 & -0.62$\pm$0.06 (0.11) & 37 & - & - \\
NGC~330       & 0.075 & -0.81$\pm$0.03 (0.10) & 220 & -0.88$\pm$0.05 (0.09) & 10 \\
              &  & -0.74$\pm$0.03 (0.10) & 247 & -0.82$\pm$0.05 (0.09) & 8 \\
OGLE-CL~SMC~45  & 0.063 & -0.72$\pm$0.02 (0.10) & 277 & -0.59$\pm$0.12 (0.09) & 3 \\
Lindsay~27    & 0.068 & -0.86$\pm$0.03 (0.10) & 181 & -0.81$\pm$0.01 (0.09) & 2 \\
Lindsay~19    & 0.062 & -0.65$\pm$0.03 (0.10) & 97 & - & - \\
NGC~265       & 0.075 & -0.62$\pm$0.03 (0.10) & 243 & -0.67$\pm$0.04 (0.09) & 31 \\
              & & -0.68$\pm$0.03 (0.10) & 202 & -0.75$\pm$0.05 (0.09) & 27 \\
NGC~376       & 0.067 & -0.62$\pm$0.03 (0.10) & 135 & -0.65$\pm$0.05 (0.09) & 7 \\
              & & -0.79$\pm$0.03 (0.10) & 170 & -0.71$\pm$0.05 (0.10) & 13 \\
IC~1611 / IC~1612  & 0.085 & -0.51$\pm$0.03 (0.10) & 148 & -0.69$\pm$0.12 (0.09) & 5 \\
OGLE-CL~SMC~32 / Bruck~39 & 0.083 & -0.52$\pm$0.03 (0.10) & 207 & -0.55$\pm$0.09 (0.09) & 6 \\
OGLE-CL~SMC~68  & 0.060 & -0.73$\pm$0.02 (0.10) & 312 & -0.77$\pm$0.05 (0.09) & 20 \\
OGLE-CL~SMC~69  & 0.074 & -0.78$\pm$0.02 (0.10) & 373 & -0.74$\pm$0.04 (0.09) & 32 \\
OGLE-CL~SMC~99  & 0.052 & -1.06$\pm$0.02 (0.10) & 394 & -1.03$\pm$0.03 (0.09) & 28 \\
OGLE-CL~SMC~129 & 0.089 & -0.69$\pm$0.02 (0.10) & 197 & -0.48$\pm$0.05 (0.11) & 12 \\
OGLE-CL~SMC~142 / [BS95]~123 / SMC~144 & 0.070 & -0.69$\pm$0.03 (0.10) & 126 & -1.17$\pm$0.26 (0.09) & 3 \\
OGLE-CL~SMC~49  & 0.065 & -0.74$\pm$0.03 (0.10) & 279 & -0.79$\pm$0.04 (0.10) & 17 \\
OGLE-CL~SMC~50  & 0.123 & -0.86$\pm$0.03 (0.10) & 346 & -0.75$\pm$0.10 (0.09) & 5 \\
OGLE-CL~SMC~54  & 0.058 & -0.89$\pm$0.03 (0.10) & 285 & -0.91$\pm$0.07 (0.10) & 16 \\
OGLE-CL~SMC~61  & 0.082 & -0.79$\pm$0.03 (0.10) & 280 & -0.63$\pm$0.04 (0.10) & 36 \\
OGLE-CL~SMC~71 / SMC~205  & 0.062 & -0.86$\pm$0.02 (0.10) & 311 & -0.90$\pm$0.04 (0.09) & 26 \\
OGLE-CL~SMC~78  & 0.046 & -0.52$\pm$0.03 (0.09) & 265 & -0.43$\pm$0.04 (0.09) & 31 \\
OGLE-CL~SMC~82 / SMC~88  & 0.057 & -0.76$\pm$0.03 (0.10) & 218 & -0.65$\pm$0.04 (0.09) & 29 \\
OGLE-CL~SMC~126 & 0.063 & -0.65$\pm$0.03 (0.09) & 203 & -0.60$\pm$0.04 (0.10) & 7 \\
OGLE-CL~SMC~128 & 0.087 & -0.35$\pm$0.03 (0.11) & 197 & -0.36$\pm$0.07 (0.10) & 3 \\
OGLE-CL~SMC~143 & 0.089 & -0.42$\pm$0.04 (0.11) & 118 & -0.26$\pm$0.03 (0.12) & 13 \\
OGLE-CL~SMC~156 & 0.068 & -0.57$\pm$0.03 (0.10) & 108 & -0.56$\pm$0.04 (0.10) & 3 \\
\hline
\end{tabular}
\end{table*}
%


%
\longtab{
\begin{landscape}
\begin{longtable}{lllll}
\caption{\label{tab:lit} Reddenings, metallicities and ages of star clusters from the literature.}\\
\hline\hline
Name & $E(B-V)$  & [Fe/H] & Age & log(Age) \\
 & (mag) & (dex) & (Gyr) & (yr) \\
\hline
\endfirsthead
\caption{continued.}\\
\hline\hline
Name & $E(B-V)$  & [Fe/H] & Age & log(Age) \\
 & (mag) & (dex) & (Gyr) & (yr) \\
\hline
\endhead
\hline
Lindsay~113     & 0.00$\pm$0.02$^5$, 0.03$\pm$0.01$^{11,15,23}$ & -1.37$\pm$0.16$^4$,
                -1.44$\pm$0.16$^4$
                & 6.0$\pm$1.0$^4$, 4.0$\pm$0.7$^5$ & 9.55$\pm$0.05$^{23}$, 9.65$^{30}$ \\
                & 0.047$^{16}$, 0.04$^{26}$, 0.02$^{27}$, 0.01$^{28}$ & -1.12$\pm$0.12$^4$,
                -1.17$\pm$0.12$^4$ & 5.3$\pm$1.3$^5$, 5.3$\pm$1.0$^11$ & \\
                & 0.0$^{22a}$, 0.077$^{32b}$ & -1.24$\pm$0.11$^4$, -1.40$\pm$0.25$^11$ &
                4.0$\pm$0.4$^{20}$, 4.6$\pm$1.0$^{26}$ & \\
                &  & -1.7$\pm$0.3$^{20}$, -1.3$\pm$0.3$^{20}$ & 4.5$\pm$0.5$^{27,28}$ & \\
                &  & -0.88$\pm$0.065$^{23}$, -1.19$\pm$0.03$^{26}$ &  & \\
                &  & -1.4$^{27}$, -1.3$^{28}$, -1.03$^{30}$ &  &  \\

\hline \\
NGC~339         & 0.03$^{1,14,27}$, 0.03$\pm$0.04$^5$, 0.032$^{12}$ & -1.5$\pm$0.2$^1$,
                -1.36$\pm$0.1$^4$ & $>$10$^1$, 4.0$\pm$1.5$^4$, 5.0$\pm$0.6$^5$ & 9.8$\pm$0.09$^{19}$,
                9.78$^{30}$ \\
                & 0.04$\pm$0.01$^{19}$, 0.020$\pm$0.007$^{23}$ & -1.46$\pm$0.1$^4$, -1.12$\pm$0.1$^4$
                & 6.3$\pm$1.3$^5$, 6.6$^{14}$, 6.0$\pm$0.5$^{18}$ & \\
                & 0.02$^{26}$, 0.054$^{32b}$ & -1.19$\pm$0.12$^{4,30}$, -1.5$\pm$0.14$^5$
                & 5.4$\pm$1.0$^{19}$ & \\
                &  & -1.18$^{14}$, -1.3$\pm$0.25$^{19}$ &  &  \\
                &  & -0.88$\pm$0.065$^{23}$, -1.27$\pm$0.03$^{26}$ &  &  \\

\hline \\
NGC~361         & 0.03$^{1,26}$, 0.07$\pm$0.03$^{5,14}$ & -1.25$\pm$0.20$^1$,
                -1.45$\pm$0.11$^{5,14,30}$
                & 8.0$\pm$1.5$^1$, 6.8$\pm$0.5$^5$ & 9.91$^{30}$ \\
                & 0.037$^{16}$, 0.0$^{22}$, 0.083$^{31}$, 0.102$^{32b}$ & -1.08$\pm$0.10$^{14,26}$,
                -0.9$\pm$0.2$^{15}$ & 8.1$\pm$1.2$^5$, 5.6$^{14}$, 6.8$\pm$0.5$^{16}$ & \\
                &  & -0.8$\pm$0.3$^{20}$, -0.7$\pm$0.3$^{20}$ & 2.0$\pm$0.4$^{20}$ & \\
                &  & -1.13$\pm$0.03$^{26}$ &  &  \\

\hline \\
Lindsay~1       & 0.06$\pm$0.02$^5$, 0.02$^{12,27}$, 0.04$^{26}$ & -1.14$\pm$0.10$^{1,12,27}$,
                -1.17$\pm$0.10$^1$ & 10.0$\pm$2.0$^1$,  7.7$\pm$0.4$^5$ & 9.88$^{30}$ \\
                & 0.044$^{32b}$ & -0.99$\pm$0.11$^1$, -1.01$\pm$0.11$^{1,14}$ & 9.0$\pm$1.0$^5$,
                8.3$\pm$0.7$^{12}$ & \\
                &  & -1.35$\pm$0.08$^5$, -1.10$\pm$0.10$^5$ & 7.7$\pm$0.7$^{12}$,
                7.5$\pm$0.5$^{12,18,26,27}$ & \\
                &  & -1.11$\pm$0.02$^{26}$, -1.04$\pm$0.03$^{30}$ &  & \\

\hline \\
Lindsay~6       & 0.03$^{19}$, 0.020$\pm$0.006$^{23}$ & -0.9$\pm$0.2$^{19}$, -1.24$\pm$0.03$^{22,30}$
                & 3.3$\pm$0.7$^{19}$, 7.0$^{22}$, 7.08$^{22}$ & 9.75$\pm$0.07$^{23}$, 9.73$^{30}$ \\
                & 0.095$^{22b}$, 0.063$^{31}$, 0.044$^{32b}$ & -0.88$\pm$0.065$^{23}$ &  & \\

\hline \\
NGC~330         & 0.03$^{3,29}$, 0.06$^{3}$, 0.10$^{8,17}$ & -1.26$^3$, -1.17$^3$, -0.93$\pm$0.16$^3$
                & 0.04$\pm$0.4$^{20}$, 0.04$^{24}$ & 7.5$\pm$0.1$^8$, 8.0$^9$, 7.4$^{17}$ \\
                & 0.11$^{24}$, 0.05$^{20a}$, 0.08$^{9b}$ & -0.69$\pm$0.11$^{7}$, -0.94$\pm$0.02$^{6}$
                &  & 7.63$^{30}$ \\
                & 0.095$^{31}$, 0.072$^{32b}$ & -1.0$\pm$0.1$^{2}$, -0.5$\pm$0.3$^{20}$, &  &  \\
                &  & -1.0$\pm$0.3$^{20}$, -0.90$^{24}$, -1.15$\pm$0.06$^{29}$, &  &  \\
                &  &  -0.82$\pm$0.10$^{30}$ &  &  \\

\hline \\
OGLE-CL~SMC~45  & 0.07$^8$, 0.02$^{13}$, 0.10$^{17}$, 0.03$^{29}$ & -0.7$^{12}$,
                -0.01$\pm$0.09$^{23,30}$ & 0.22$^{13}$ & 8.4$\pm$0.1$^{8,9}$, 8.35$^{17}$, 8.38$^{30}$ \\
                & 0.10$\pm$0.04$^{23}$, 0.08$^{9b}$, 0.077$^{31}$, 0.064$^{32b}$ & -0.85$\pm$0.15$^{29}$
                &  & 6.9$\pm$0.07$^{23}$ \\

\hline \\
Lindsay~27      & 0.06$^8$, 0.11$^{15,19}$, 0.00$\pm$0.03$^{23}$ & -1.3$\pm$0.3$^{19}$,
                -1.14$\pm$0.06$^{22,30}$ & 2.1$\pm$0.3$^{11}$, 6.3$\pm$1.0$^{22}$ & $>$9$^8$,
                9.45$\pm$0.06$^{23}$, 9.66$^{30}$ \\
                & 0.084$^{31}$, 0.067$^{32b}$ & -0.88$\pm$0.065$^{23}$ & 4.6$\pm$0.6$^{22}$ &  \\

\hline \\
Lindsay~19      & 0.10$^8$, 0.02$^{15,19}$, 0.04$^{9b,23}$  & -0.75$\pm$0.20$^{11}$,
                -0.87$\pm$0.03$^{22,30}$ & 2.1$\pm$0.3$^{11,19}$ & $>$9$^8$, 8.9$^9$,
                9.45$\pm$0.08$^{23}$ \\
                & 0.16$^{25b}$, 0.079$^{31}$, 0.058$^{32b}$ & -0.88$\pm$0.065$^{23}$ &  & 8.5$^{25}$,
                9.6$^{30}$ \\

\hline \\
NGC~265         & 0.11$^8$, 0.05$^{11}$, 0.09$^{17}$ & -0.62$^9$, -0.6~--~-1.1$^{11}$
                & 0.25$\pm$0.12$^{11}$ & 8.0$\pm0.1^8$, 8.4$^9$, 8.5$\pm$0.3$^{10}$ \\
                & 0.04$\pm$0.02$^{23}$, 0.08$^{9b}$, 0.17$^{25b}$ & -0.28$\pm$0.16$^{23,30}$ &  &
                8.7$\pm$0.2$^{10}$, 8.35$^{17}$, 8.25$^{25,30}$ \\ 
                & 0.077$^{31}$, 0.095$^{32b}$ &  &  & 8.45$\pm$0.06$^{23}$ \\

\hline \\
NGC~376         & 0.07$^8$, 0.14$^{11}$, 0.08$^{17}$, 0.03$^{29}$ & -0.5$\pm$0.3$^{20}$,
                -0.40$\pm$0.29$^{23,30}$ & 0.025$\pm$0.01$^{11}$, 0.028$^{29}$ & 7.5$\pm$0.1$^{8,17}$,
                7.2$^9$, 7.8$^{23}$ \\
                & 0.00$\pm$0.02$^{23}$, 0.4$^{20}$, 0.08$^{9b}$ & -0.55$\pm$0.09$^{29}$
                & 0.03 $\pm$ 0.4$^{20}$ & 7.35$^{25}$, 7.4$^{30}$ \\
                & 0.15$^{25b}$, 0.090$^{31}$, 0.059$^{32b}$ &  &  & \\

\hline \\
IC~1611         & 0.08$^8$, 0.15$^{11}$, 0.05$^{17}$, 0.03$^{29}$ & -0.7$^{29}$,
                -0.80$\pm$0.09$^{29}$ &
                0.10$\pm$0.04$^{29}$ & 8.2$\pm$0.1$^{8,17}$, 8.1$^9$, 8.15$^{25}$ \\
                & 0.08$^{9b}$, 0.06$^{25b}$, 0.110$^{31}$, 0.079$^{32b}$ &  &  & \\

\hline \\
IC~1612         & 0.07$^8$, 0.08$^{9b}$, 0.06$^{13}$, 0.13$^{17}$ & -0.7$^{13}$,
                -0.10$\pm$0.11$^{23,30}$ &  & 7.7$\pm$0.2$^{8}$, 8.0$^{9,17}$, 8.09$^{13}$ \\
                & 0.05$\pm$0.01$^{23}$, 0.21$^{25b}$, 0.110$^{31}$, 0.079$^{32b}$ &  & - &
                8.10$\pm$0.05$^{23}$, 7.47$^{25}$, 7.90$^{30}$ \\
                & 0.079$^{32b}$ &  &  &  \\

\hline \\
OGLE-CL~SMC~32  & 0.1$^8$, 0.05$^{11}$, 0.2$^{17}$ & -0.10$\pm$0.18$^{23,30}$ & 0.16$\pm$0.07$^{11}$
                & 8.0$\pm$0.1$^8$, 8.0$^{9,23}$, 7.8$^{17,25}$ \\
                & 0.06$\pm$0.03$^{23}$, 0.2$^{9b}$, 0.26$^{25b}$ &  &  & 7.93$^{30}$ \\
                & 0.103$^{31}$, 0.083$^{32b}$ &  &  & \\

\hline \\
Bruck~39        & 0.08$^{9b}$, 0.08$^{17}$, 0.02$\pm$0.04$^{23}$ & -0.58$\pm$0.54$^{23,30}$ & - &
                8.6$\pm$0.2$^{9}$, 8.65$^{17}$, 8.9$^{23}$, 8.3$^{25}$ \\
                & 0.20$^{25b}$, 0.103$^{31}$, 0.083$^{32b}$ &  &  & 9.08$^{30}$ \\

\hline \\
OGLE-CL~SMC~68  & 0.09$^8$, 0.25$^{17}$, 0.16$^{9b}$, 0.077$^{31}$ & - & - & 7.7$\pm$0.2$^8$, 7.0$^9$,
                7.4$^{17}$ \\
                & 0.057$^{32b}$ &  &  & 7.37$^{30}$ \\

\hline \\
OGLE-CL~SMC~69  & 0.08$^8$, 0.1$^{17}$, 0.2$^{9b}$, 0.3$^{25b}$ & -0.75$^{10}$ & - & 7.6$\pm$0.1$^8$,
                7.0$^9$,  \\
                & 0.088$^{31}$, 0.080$^{32b}$ &  &  & 7.8$\pm$0.5$^{10}$, 8.0$\pm$0.3$^{10}$,  \\
                &  &  &  & 7.6$^{17}$, 7.63$^{25}$, 7.4$^{30}$ \\

\hline \\
OGLE-CL~SMC~99  & 0.08$^8$, 0.05$^{17}$, 0.08$^{9b}$ & - & - & 7.6$\pm$0.2$^8$, 7.3$^9$, 7.5$^{17}$ \\
                & 0.11$^{25b}$, 0.069$^{31}$, 0.045$^{32b}$ &  &  & 7.7$^{25}$, 7.47$^{30}$ \\

\hline \\
OGLE-CL~SMC~129 & 0.09$^8$, 0.05$^{17}$, 0.12$^{9b}$ & - & - & 7.3$\pm$0.1$^8$, 7.3$^9$, 7.4$^{17}$ \\
                & 0.098$^{31}$, 0.105$^{32b}$ &  &  & 7.33$^{30}$ \\

\hline \\
OGLE-CL~SMC~142 & 0.06$^8$, 0.08$^{17}$, 0.14$^{9b}$, 0.13$^{25b}$ & - & - & 7.3$\pm$0.1$^8$, 6.9$^9$,
                7.0$^{17}$ \\
                & 0.078$^{31}$, 0.082$^{32b}$ &  &  & 7.24$^{25}$, 7.07$^{30}$ \\

\hline \\
$[\mathrm{BS95}]$~123      & 0.08$^{9b}$, 0.07$^{17}$, 0.11$^{25b}$, 0.078$^{31}$ & - & -
                & 7.5$^{9}$, 7.6$^{17}$, 7.95$^{25}$, 7.55$^{30}$ \\
                & 0.082$^{32b}$ &  &  &  \\

\hline \\
OGLE-CL~SMC~144 & 0.05$^{8}$, 0.10$^{9b}$, 0.05$^{17}$, 0.078$^{31}$ & - & -
                & 7.6$^{8,9}$, 7.7$^{17}$, 7.67$^{30}$ \\
                & 0.082$^{32b}$ &  &  &  \\

\hline \\
OGLE-CL~SMC~49  & 0.06$^8$, 0.02$^{17}$, 0.06$\pm$0.03$^{23}$ & -0.10$\pm$0.14$^{23,30}$ & - &
                7.0$\pm$0.2$^8$, 8.0$^9$, 7.9$^{17}$ \\
                & 0.08$^{9b}$, 0.086$^{31}$, 0.058$^{32b}$ &  &  & 7.5 $\pm$ 0.3$^{23}$, 7.7$^{30}$ \\

\hline \\
OGLE-CL~SMC~50  & 0.12$^8$, 0.16$^{17}$, 0.15$^{9b}$ & - & - & 8.1$\pm$0.1$^8$, 7.9$^9$,
                8.2$^{17}$ \\
                & 0.11$^{31}$, 0.197$^{32b}$ &  &  & 8.0$^{30}$ \\

\hline \\
OGLE-CL~SMC~54  & 0.10$^8$, 0.08$^{17}$, 0.05$^{21}$, 0.04$\pm$0.03$^{23}$ & -0.01$\pm$0.09$^{23,30}$
                & - & 8.0$\pm$0.1$^8$, 8.0$^{9,17}$, 8.01$^{30}$ \\
                & 0.08$^{9b}$, 0.13$^{25b}$, 0.079$^{31}$, 0.048$^{32b}$ &  &  & 7.0$\pm$0.4$^{23}$,
                8.25$^{25}$ \\

\hline \\
OGLE-CL~SMC~61  & 0.12$^8$, 0.15$^{9b}$, 0.104$^{31}$, 0.081$^{32b}$ & - & - & 7.4$\pm$0.2$^8$,
                7.8$^9$, 7.6$^{30}$ \\

\hline \\
OGLE-CL~SMC~71  & 0.07$^8$, 0.06$^{17}$, 0.2$^{9b}$, 0.34$^{25b}$ & - & - & 7.5$\pm$0.1$^8$,
                7.1$^9$, 8.75$^{17}$ \\
                & 0.087$^{31}$, 0.049$^{32b}$ &  &  & 7.88$^{25}$, 8.2$^{30}$ \\

\hline \\
OGLE-CL~SMC~205 & 0.08$^{9b,17}$, 0.087$^{31}$, 0.049$^{32b}$ & - & -
                & 8.1$^{9}$, 8.3$^{17}$, 8.2$^{30}$ \\

\hline \\
OGLE-CL~SMC~78  & 0.08$^{8,17}$, 0.08$^{9b}$, 0.074$^{31}$ & - & - & 7.9$\pm$0.1$^8$, 7.0$^9$,
                7.8$^{17}$ \\
                & 0.025$^{32b}$ &  &  & 7.57$^{30}$ \\

\hline \\
OGLE-CL~SMC~82  & 0.10$^{8,17}$, 0.15$^{9b}$, 0.085$^{31}$ & - & - & 7.8$\pm$0.3$^8$, 7.9$^{9,17}$,
                7.87$^{30}$ \\
                & 0.038$^{32b}$ &  &  &  \\

\hline \\
OGLE-CL~SMC~88  & 0.20$^{9b}$, 0.085$^{31}$, 0.038$^{32b}$ & - & - & 7.2$^{9,30}$ \\

\hline \\
 OGLE-CL~SMC~126 & 0.07$^8$, 0.02$^{17}$, 0.08$^{9b}$ & - & - & 8.0$\pm$0.1$^8$, 8.0$^9$, 8.2$^{17}$ \\
                 & 0.17$^{25b}$, 0.083$^{31}$, 0.057$^{32b}$  &  &  & 7.89$^{25}$, 8.07$^{30}$ \\

\hline \\
OGLE-CL~SMC~128  & 0.09$^8$, 0.07$^{17}$, 0.08$^{9b}$ & - & - & 7.1$\pm$0.3$^8$, 7.3$^9$,
                 8.0$^{17}$ \\
                 & 0.099$^{31}$, 0.098$^{32b}$ &  &  & 7.2$^{30}$ \\

\hline \\
OGLE-CL~SMC~143  & 0.09$^8$, 0.2$^{9b}$, 0.093$^{31}$, 0.113$^{32b}$ & - & - & 8.2$\pm$0.2$^8$,
                 7.5$^9$, 7.85$^{30}$ \\

\hline \\
OGLE-CL~SMC~156  & 0.09$^8$, 0.10$^{17}$, 0.00$^{20a}$, 0.08$^{9b}$ & -1.0$\pm$0.3$^{20}$,
                  -0.7$\pm$0.3$^{20}$ & 0.2$\pm$0.4$^{20}$ & 8.2$\pm$0.1$^8$, 8.1$^{9,17,23}$  \\
                 & 0.2$^{25b}$, 0.088$^{31}$, 0.064$^{32b}$ &  &  & 7.8$^{25}$ \\
\hline
\end{longtable}
\tablefoot{
  \begin{itemize}\scriptsize
  \itemsep0em
    \item [$^{(a)}$] E(b-y).
    \item [$^{(b)}$] E(V-I).
    \item[*] References: (1) \citet{Bica1986} (DDO, H$\beta$ photometry);
    (2) \citet{Spite1991} (high-resolution spectroscopy);
    (3) \citet{GR1992} (Str\"omgren photometry);
    (4) \citet{DH1998} (low-resolution spectroscopy, CaT);
    (5) \citet{Mighell1998} (HST photometry, RGB slope);
    (6) \citet{GW1999} (high-resolution spectroscopy);
    (7) \citet{Hill1997} (high-resolution spectroscopy);
    (8) \citet{PU1999a} (optical photometry);
    (9) \citet{Chiosi2006} (optical photometry);
    (10) \citet{CV2007} (HST photometry);
    (11) \citet{Piatti2007a,Piatti2007b} (Washington photometry);
    (12) \citet{Glatt2008} (HST photometry);
    (13) \citet{Piatti2008} (Washington photometry);
    (14) \citet{Mucciarelli2009} (NIR photometry, RGB slope);
    (15) \citet{Kucinskas2009} (NIR photometry, RGB slope);
    (16) \citet{Dias2010} (integrated spectroscopy);
    (17) \citet{Glatt2010} (optical photometry);
    (18) \citet{Glatt2011} (HST photometry);
    (19) \citet{Piatti2011} (Washington photometry);
    (20) \citet{Livanou2013} (Str\"omgren photometry);
    (21) \citet{Maia2014} (Washington photometry);
    (22) \citet{Parisi2014} (low-resolution spectroscopy, CaT);
    (23) \citet{Perren2017} (Washington photometry, ASteCA);
    (24) \citet{Milone2018} (HST photometry);
    (25) \citet{Nayak2018} (optical photometry);
    (26) \citet{Piatti2018} (Str\"omgren photometry);
    (27) \citet{Chantereau2019} (HST photometry);
    (28) \citet{Martocchia2019} (HST photometry);
    (29) \citet{Piatti2019} (Str\"omgren photometry);
    (30) \citet{Bica2020} (catalog);
    (31) \citet{Gorski2020} (optical photometry);
    (32) \citet{Skowron2020} (optical photometry).
  \end{itemize}
}
\end{landscape}
}
%

\begin{table*}
\caption{Str\"omgren photometry of fields in the SMC.}
\label{tab:phot}
\centering
\begin{tabular}{ccrrcrccrcc}
\hline\hline
RA & DEC & X & Y & Field
& $V$ & $\sigma_{V_{DAO}}$ & $\sigma_{V}$
& $(b-y)$ & $\sigma_{(b-y)_{DAO}}$ & $\sigma_{(b-y)}$ \\
(deg) & (deg) & (pixel) & (pixel) & &
(mag) & (mag) & (mag) & (mag) & (mag) & (mag) \\
\hline
27.531676 & -73.741133 &  2.534 &  666.291 & Lindsay113 & 20.324 & 0.021 & 0.027
& 0.326 & 0.039 & 0.042 \\
27.530038 & -73.752099 & 12.939 &  408.817 & Lindsay113 & 21.318 & 0.045 & 0.049
& 0.439 & 0.064 & 0.067 \\
27.528188 & -73.712040 & 26.633 & 1349.382 & Lindsay113 & 14.134 & 0.006 & 0.015
& -0.050 & 0.009 & 0.014 \\
27.527827 & -73.758950 & 27.267 &  247.924 & Lindsay113 & 22.112 & 0.092 & 0.093
& 0.081 & 0.113 & 0.115 \\
 \multicolumn{11}{l}{(...)}\\ 
\hline
\end{tabular}
\begin{tabular}{rccrr}
\hline\hline
$m1$ & $\sigma_{m1_{DAO}}$ & $\sigma_{m1}$ & CHI & SHARP  \\
(mag) & (mag) & (mag) & & \\
\hline
0.171 & 0.161 & 0.178 & 0.727 & 0.055 \\
-0.116 & 0.117 & 0.132 & 0.763 & -0.067 \\
0.060 & 0.016 & 0.024 & 3.371 & 0.351 \\
0.408 & 0.174 & 0.194 & 0.807 & -0.724 \\
 \multicolumn{5}{l}{(...)}\\ 
\hline
\end{tabular}
\tablefoot{A~complete table is presented in its entirety in the electronic
form on the Araucaria Project webpage and the CDS. A~portion is shown
here for guidance regarding its form and content.}
\end{table*}

%
%

\begin{appendix} 
\section{Comments on individual clusters}
There were a~few cases when the reddening value adopted for a~given star cluster
as a~mean of G20 and S21 was unsuitable for age determination. The isochrones
for calculated metallicities were not fitting well the observed CMDs.
In case of Lindsay~113, $E(B-V)_{GS} = 0.058$~mag from S21 turned out to be too
high. The isochrone of $\mathrm{[Fe/H] = -1.01}$~dex was not fitting well
simultaneously the subgiant and red giant branch, suggesting that a~steeper
isochrone for lower metallicity is needed. To that end, we adopted
$E(B-V) = 0.03$~mag often used in the literature. Similar situation was in case
of NGC~361 and Lindsay~1 having $E(B-V)_{GS} = 0.08$~mag and 0.049~mag,
respectively, for which also $E(B-V) = 0.03$~mag was adopted. Also in case of
IC~1611 having $E(B-V)_{GS} = 0.085$~mag the isochrone of $-0.19$~dex did not
fit well the main sequence. We adopted $E(B-V) = 0.03$~mag instead which
resulted in $\mathrm{[Fe/H]} = -0.58$~dex.

\citet{Perren2017} reported $\mathrm{[Fe/H] = -0.01}$~dex for OGLE-CL~SMC~45
while \citet{Piatti2019} give $-0.85$~dex for $E(B-V) = 0.03$~mag. For
$E(B-V)_{GS} = 0.063$~mag we got $\mathrm{[Fe/H] = -0.31}$~dex.
The differences between this work and \citet{Piatti2019} using the same data
could be caused by the choice of different stars for the mean metallicity
calculation, photometric and calibration errors, as well as the use of different
two-color calibration.

NGC~376 is too metal-poor compared to the literature. This cluster was observed
twice during two subsequent nights. For its metallicity calculation we have
chosen the same two stars as used in Piatti et al. (2019). The reported mean metallicity
of the cluster is $-0.55$~dex which is much higher value than what we obtained
in this work ($-0.98$~dex). One of the stars is lying in the very center of the cluster
and its photometry could be affected by the dense environment. Its metallicity agrees
well between the two nights though ($-1.20\pm0.28$~dex and $-1.18\pm0.27$~dex in the
second and third night, respectively). The other chosen star was measured only in the
image from the third night, because on the second it fell into the gap between the
chips. We obtained for it $\mathrm{[Fe/H]} = -0.77\pm0.25$, which is also more
metal-poor than in Piatti et al. (2019).
The isochrone of $-0.77$~dex seems to fit well the CMD of NGC~376, suggesting that it
could be more metal-rich and simultaneously younger ($\mathrm{log(Age_P)} \approx
7.50$~yr) than the age of best-fitting isochrone of $-0.98$~dex. Livanou et al. (2013)
have only one star in this region of the CMD and in their case it is very metal-poor.

The S21 reddening map gives very small reddening values for OGLE-CL~SMC~78 and 82,
very different than G20 and other authors. We used the average of S21 and G20
for metallicity calculations, but higher reddening values resulting in higher
metallicities and ages would be equally acceptable for these clusters.

\end{appendix}

\end{document}